\documentstyle[epsfig,a4p,12pt]{article}

\begin{document}
%
%
\newcommand{\PPEnum}    {CERN-PPE/97-083}
\newcommand{\PNnum}     {OPAL Physics Note PN278}
\newcommand{\Date}      {14th June 1997}
\newcommand{\Author}    {S.~Asai, A.~F\"{u}rtjes, M.~Gruw\'e, J.~Kanzaki,
S.~Komamiya, D.~Lellouch, G.~Mikenberg, H.~Neal, and R.~Van Kooten}
\newcommand{\MailAddr}  {Magali.Gruwe@cern.ch}
\newcommand{\EdBoard}   {R.~Barlow, H.~Evans, S.~de~Jong and P.~Krieger} 
\newcommand{\DraftVer}  {Version 1.00}
\newcommand{\DraftDate} {\today}
\newcommand{\TimeLimit} {03. July 1997 18:00 CERN time} 
\def\toprule{\noalign{\hrule \medskip}}
\def\midrule{\noalign{\medskip\hrule }}
\def\botrule{\noalign{\medskip\hrule }}
\setlength{\parskip}{\medskipamount}
 

\newcommand{\lumi}{10.3}
\newcommand{\ee}{{\mathrm e}^+ {\mathrm e}^-}
\newcommand{\sq}{\tilde{\mathrm q}}
\newcommand{\sg}{\tilde{\mathrm g}}
\newcommand{\su}{\tilde{\mathrm u}}
\newcommand{\sd}{\tilde{\mathrm d}}
\newcommand{\seff}{\tilde{\mathrm f}}
\newcommand{\sele}{\tilde{\mathrm e}}
\newcommand{\sell}{\tilde{\ell}}
\newcommand{\snu}{\tilde{\nu}}
\newcommand{\smu}{\tilde{\mu}}
\newcommand{\ch}{\tilde{\chi}^\pm}
\newcommand{\chp}{\tilde{\chi}^+}
\newcommand{\chm}{\tilde{\chi}^-}
\newcommand{\chpm}{\tilde{\chi}^\pm}
\newcommand{\chmp}{\tilde{\chi}^\mp}
\newcommand{\nt}{\tilde{\chi}^0}
\newcommand{\qq}{{\mathrm q}\bar{\mathrm q}}
\newcommand{\WW}{{\mathrm W}^+ {\mathrm W}^-}
\newcommand{\gv}{\gamma^*}
\newcommand{\Wenu}{{\mathrm W e}\nu}
\newcommand{\ZZ}{{\mathrm Z} {\mathrm Z}}
\newcommand{\Zg}{{\mathrm Z} \gamma}
\newcommand{\Zgv}{{\mathrm Z} \gamma^*}
\newcommand{\Zvgv}{{\mathrm Z^*} \gamma^*}
\newcommand{\Wv}{{\mathrm W}^{*}}
\newcommand{\Wrv}{{\mathrm W}^{(*)}}
\newcommand{\Wvp}{{\mathrm W}^{*+}}
\newcommand{\Wvm}{{\mathrm W}^{*-}}
\newcommand{\Wvpm}{{\mathrm W}^{*\pm}}
\newcommand{\Z}{\mathrm Z}
\newcommand{\Zv}{{\mathrm Z}^{*}}
\newcommand{\Zrv}{{\mathrm Z}^{(*)}}
\newcommand{\ffbar}{{\mathrm f}\bar{\mathrm f}}
\newcommand{\nunu}{\nu \bar{\nu}}
\newcommand{\mumu}{\mu^+ \mu^-}
\newcommand{\tautau}{\tau^+ \tau^-}
\newcommand{\ellell}{\ell^+ \ell^-}
\newcommand{\nulqq}{\nu \ell {\mathrm q} \bar{\mathrm q}'}
\newcommand{\MZ}{m_{\mathrm Z}}
 
\newcommand{\gsim}{\;\raisebox{-0.9ex}
           {$\textstyle\stackrel{\textstyle >}{\sim}$}\;}
\newcommand{\lsim}{\;\raisebox{-0.9ex}{$\textstyle\stackrel{\textstyle<}
           {\sim}$}\;}

\newcommand{\degree}    {^\circ}
%
\newcommand{\Ecm}       {E_{\mathrm{cm}}}
\newcommand{\Ebeam}     {E_{\mathrm{b}}}
\newcommand{\roots}     {\sqrt{s}}
%
%
\newcommand{\thrust}    {T}
\newcommand{\nthrust}   {\hat{n}_{\mathrm{thrust}}}
\newcommand{\thethr}    {\theta_{\,\mathrm{thrust}}}
\newcommand{\phithr}    {\phi_{\mathrm{thrust}}}
\newcommand{\acosthr}   {|\cos\thethr|}
\newcommand{\thejet}    {\theta_{\,\mathrm{jet}}}
\newcommand{\acosjet}   {|\cos\thejet|}
\newcommand{\thmiss}    { \theta_{\mathrm{miss}} }
\newcommand{\cosmiss}   {| \cos \thmiss |}
%
%
\newcommand{\Evis}      {E_{\mathrm{vis}}}
\newcommand{\Rvis}      {E_{\mathrm{vis}}\,/\roots}
\newcommand{\Mvis}      {M_{\mathrm{vis}}}
\newcommand{\Rbal}      {R_{\mathrm{bal}}}
\newcommand{\pt}        {p_{\mathrm{t}}}
%
%
\newcommand{\phiacop}   {\phi_{\mathrm{acop}}}
\newcommand{\axicos}{{\mid\cos\theta_a}^{\mathrm{miss}}\mid}
%
%
\newcommand{\PhysLett}  {Phys.~Lett.}
\newcommand{\PRL} {Phys.~Rev.\ Lett.}
\newcommand{\PhysRep}   {Phys.~Rep.}
\newcommand{\PhysRev}   {Phys.~Rev.}
\newcommand{\NPhys}  {Nucl.~Phys.}
\newcommand{\NIM} {Nucl.~Instr.\ Meth.}
\newcommand{\CPC} {Comp.~Phys.\ Comm.}
\newcommand{\ZPhys}  {Z.~Phys.}
\newcommand{\IEEENS} {IEEE Trans.\ Nucl.~Sci.}
%
%
\newcommand{\OPALColl}  {OPAL Collab.}
\newcommand{\JADEColl}  {JADE Collab.}
\newcommand{\etal}      {{\it et~al.}}
\newcommand{\onecol}[2] {\multicolumn{1}{#1}{#2}}
\newcommand{\ra}        {\rightarrow}   
 

 
\begin{titlepage}
%
%
\begin{center}
    \large
    EUROPEAN LABORATORY FOR PARTICLE PHYSICS
\end{center}
\begin{flushright}
    \large
   \PPEnum\\
  \Date
\end{flushright}
%
%
%
%
\begin{center}
    \Large\bf\boldmath
    Search for Chargino and Neutralino Production 
    at $\sqrt{s} = 170$ and 172~GeV at LEP
\end{center}
%
%
\begin{center}
    \Large
    The OPAL Collaboration \\
\bigskip
\bigskip
%
\end{center}
%
%
\begin{abstract}
A search for charginos and neutralinos,
predicted by supersymmetric theories, has been
performed using a data sample of $\lumi$~pb$^{-1}$
at centre-of-mass energies of $\roots = $170 and 172~GeV
with the OPAL detector
at LEP. No evidence for these particles has been found.
The results are combined with those from previous OPAL
chargino and neutralino searches at lower energies to obtain limits.
Exclusion regions at 95\% C.L. of parameters of the
Minimal Supersymmetric Standard Model are determined.
Within this framework, for $\tan\beta \ge 1.0$, lower mass limits
are placed on the lightest chargino and the three lightest neutralinos. 
The 95\% C.L. lower mass limit on the lightest chargino, assuming that
it is heavier than the lightest neutralino by
more than 10~GeV, is 84.5~GeV for the case of a large universal
scalar mass ($m_0 >$~1~TeV) and 65.7~GeV for the smallest
$m_0$ compatible with current limits on the sneutrino and slepton masses.
The lower limit on the lightest neutralino mass at
95\% C.L. for $\tan \beta \geq 1.0$ is 24.7~GeV for $m_0 = 1$~TeV
and
13.3~GeV for the minimum $m_0$ scenario.
These mass limits are higher for increasing values of $\tan\beta$.
The interpretation of the limits
in terms of gluino and scalar quark mass limits is also given.
\end{abstract}
\begin{center}
\noindent
%

{\bf \Large Submitted to Zeit. Phys. C}\\

\end{center}

\end{titlepage}
 
\begin{center}{\Large        The OPAL Collaboration
}\end{center}\bigskip
\begin{center}{
K.\thinspace Ackerstaff$^{  8}$,
G.\thinspace Alexander$^{ 23}$,
J.\thinspace Allison$^{ 16}$,
N.\thinspace Altekamp$^{  5}$,
K.J.\thinspace Anderson$^{  9}$,
S.\thinspace Anderson$^{ 12}$,
S.\thinspace Arcelli$^{  2}$,
S.\thinspace Asai$^{ 24}$,
D.\thinspace Axen$^{ 29}$,
G.\thinspace Azuelos$^{ 18,  a}$
% ...
}\end{center}\bigskip
\bigskip
$^{  1}$School of Physics and Space Research, University of Birmingham,
Birmingham B15 2TT, UK
\newline
$^{  2}$Dipartimento di Fisica dell' Universit\`a di Bologna and INFN,
I-40126 Bologna, Italy
\newline
% ...
$^{  5}$Cavendish Laboratory, Cambridge CB3 0HE, UK
\newline
% ...
$^{  8}$CERN, European Organisation for Particle Physics,
CH-1211 Geneva 23, Switzerland
\newline
$^{  9}$Enrico Fermi Institute and Department of Physics,
University of Chicago, Chicago IL 60637, USA
\newline
% ...
$^{ 12}$Indiana University, Department of Physics,
Swain Hall West 117, Bloomington IN 47405, USA
\newline
% ...
$^{ 16}$Department of Physics, Schuster Laboratory, The University,
Manchester M13 9PL, UK
\newline
% ...
$^{ 18}$Laboratoire de Physique Nucl\'eaire, Universit\'e de Montr\'eal,
Montr\'eal, Quebec H3C 3J7, Canada
\newline
% ...
$^{ 23}$Department of Physics and Astronomy, Tel Aviv University,
Tel Aviv 69978, Israel
\newline
$^{ 24}$International Centre for Elementary Particle Physics and
Department of Physics, University of Tokyo, Tokyo 113, and
Kobe University, Kobe 657, Japan
\newline
% ...
$^{ 29}$University of British Columbia, Department of Physics,
Vancouver BC V6T 1Z1, Canada
\newline
% ...
$^{  a}$ and at TRIUMF, Vancouver, Canada V6T 2A3
\newline
% ...
\newline

\newpage

 
\section{Introduction}
When unification of the
electromagnetic, weak, and strong forces in the Standard Model is
considered, a severe problem exists in the understanding of the enormous ratio
between the energy scales of full unification of the three forces and the scale
of unification of the electromagnetic and weak forces.  This ``naturalness" or
``gauge hierarchy" problem must be addressed by any new theory that attempts to
unify the fundamental forces.  One of the most promising candidates for this
new physics is the theory of supersymmetry (SUSY)~\cite{SUSY}
that extends the Standard
Model with a new type of symmetry between fermions and bosons that
also allows for the incorporation of the gravitational force.

In October and November 1996
the LEP $\ee$ collider at CERN was run
at the new centre-of-mass energies ($\roots$) of 170 and 172~GeV,
thus expanding the kinematically accessible region for new
particle searches.
A direct search for charginos
and neutralinos predicted in SUSY theories~\cite{SUSY}
is described
using the data collected with the OPAL detector.
The large predicted cross section for chargino production leads to
excellent discovery potential for this SUSY particle with the present 
integrated luminosity collected at LEP2.
More stringent exclusion mass and cross-section limits are obtained
compared to the previous results from the analysis of data
near the Z peak (LEP1),
at $\roots=130\mathrm~GeV$ and $136\mathrm~GeV$ (LEP1.5) and
at $\roots=161\mathrm~GeV$
by the OPAL~\cite{LEP15-opal,LEP16-opal} and the 
other LEP collaborations~\cite{LEP15-chargino}.
The search for charginos is similar to that described
in \cite{LEP16-opal} while a new and more efficient method has been
incorporated for the neutralino search.
Similar, but more model-dependent, limits have been obtained
by the CDF and D0 collaborations at the
Tevatron $\mathrm p\bar{p}$ collider~\cite{CDFD0-chargino}.

Charginos, $\chpm_{j}$, are the mass eigenstates
formed by the mixing of
the fields of the fermionic partners of the W boson (winos)
and those of the charged Higgs bosons (charged higgsinos).
Fermionic partners of the $\gamma$, the Z boson,
and the neutral Higgs bosons
mix to form mass eigenstates called neutralinos, $\nt_{i}$.
In each case, the index $j$
or $i$ is ordered by increasing mass.
R-parity~\cite{rpv} conservation is assumed;
therefore, the lightest supersymmetric particle (LSP)
is stable. The LSP is usually considered to be 
the lightest neutralino, $\nt_1$,
although it could be the scalar
neutrino, $\tilde{\nu}$, if it is sufficiently light.
It escapes detection due to its weakly interacting nature.
The present analysis is valid in either case.

The Minimal Supersymmetric Standard Model (MSSM)~\cite{MSSM}
is used to guide the analysis but more
general cases are also studied.
In the MSSM two chargino mass eigenstates ($\ch_1$ and $\ch_2$) 
and four neutralino mass eigenstates ($\nt_1$, $\nt_2$, $\nt_3$ and $\nt_4$)
are expected to exist.

If charginos exist and are sufficiently light,
they will be pair-produced in $\ee$ collisions
through a $\gamma$ or Z in the $s$-channel.
For the wino component there is an additional production process 
through scalar electron neutrino (electron sneutrino, $\snu_{\mathrm{e}}$) 
exchange in the $t$-channel.
The production cross-section is large
unless the sneutrino is light, in which case
the cross-section is reduced by destructive interference
between the $s$-channel $\mathrm e^+e^-$ annihilation to
Z or $\gamma$ and $t$-channel $\snu_{\mathrm{e}}$ 
exchange diagrams~\cite{Bartl, chargino-theory}.
The details of chargino decay depend on
the parameters of the mixing
and the masses of the scalar partners of the ordinary fermions.
The lightest chargino $\chp_1$ can decay
into $\nt_1 \ell^+ \nu$, or
$\nt_1 \mathrm{q} \mathrm{\overline{q}'}$,
via a virtual W,
scalar lepton (slepton, $\sell$), sneutrino ($\snu$)
or scalar quark (squark, $\sq$).
In much of the MSSM parameter space $\chp_1$
decays via a virtual W are dominant. In the MSSM,
the predicted cross-section for $\chp_1 \chm_1$ pair production is
typically several pb.
Due to the energy and momentum carried away
by the invisible $\nt_1$ and/or neutrinos,
the experimental signature for $\chp_1 \chm_1$ events
is large missing energy and large missing momentum transverse
to the beam axis.

Neutralino pairs ($\nt_1 \nt_2$) can be produced through an
$s$-channel virtual Z or $\gamma$, or by $t$-channel scalar 
electron (selectron, $\sele$)
exchange \cite{Ambrosanio}.
The $\nt_2$ will decay into the final states\footnote{The decay would
be via a Z$^*$,
a neutral SUSY Higgs boson ($\mathrm{h}^0$ or $\mathrm{A}^0$),
sneutrino, slepton, or squark.
The decay via Z$^*$ is the dominant mode in most of 
the parameter space.}
$\nt_1 \nunu$, $\nt_1 \ellell$ or
$\nt_1 {\mathrm q}\bar{\mathrm q}$.
For the latter two cases
this leads to an experimental signature consisting either
of an acoplanar pair
of particles or jets, or a monojet in which the two jets in
the final state have merged.
In some regions of SUSY parameter space the radiative decay process
$\nt_2 \ra \nt_1 \gamma$ is also possible~\cite{raddecay}.
The MSSM predicted
cross-sections
for $\nt_2 \nt_1$ events
can vary significantly
depending on the choice of MSSM parameters.
They are typically a fraction of a pb and generally much less than
the cross-section for $\chm_1 \chp_1$ production.
In the MSSM analyses reported here all possible
cascade decay processes \cite{Ambrosanio,Carena} are taken into account. 
For example,
$\nt_3$ decays into
$\nt_{1,2} \mathrm{Z}^*$
are considered as well as
$\nt_3$ decays into $\nt_{1,2}
\mathrm{h}^0$ and $\nt_{1,2} \mathrm{A}^0$
and decays where $\nt_3 \to \nt_2 \gamma$ with
$\nt_2 \to \nt_1 \mathrm{Z^*}$.
Therefore, the experimental signatures for $\nt_1 \nt_3$ events are
somewhat similar to those for $\nt_1 \nt_2$. $\nt_2 \nt_2$ pairs can
also be
produced, but as limits on their production do not affect the limits placed
on the MSSM parameter space (CMSSM, see below) they are not considered
in this analysis.

Within the framework of the MSSM
the mass spectra and couplings of char\-gi\-nos and neutralinos 
are mainly determined by the following four parameters:
the ratio of the vacuum expectation values 
of the two Higgs doublets ($\tan \beta$),
the U(1) and SU(2) gaugino mass parameters at the weak scale ($M_1$
and $M_2$), 
and the mixing parameter of the two Higgs doublet fields ($\mu$).
Assuming Grand Unification (GUT), all gauginos (supersymmetric 
partners of the gauge bosons) have a common mass $m_{1/2}$ at 
the GUT mass scale.
The gaugino masses at the weak scale are determined by the 
renormalization group equations.
As a result, the ratios of the U(1), SU(2)
and SU(3) gaugino masses ($M_1$:$M_2$:$M_3$)
are equal to $\alpha_1$:$\alpha_2$:$\alpha_3$, where 
$\alpha_i$ $(i=1,2,3)$ are the strengths of the gauge couplings 
at the weak scale.  
If Grand Unification is assumed then $M_2 = 0.82~m_{1/2}$.
$M_2$ is conventionally chosen as an independent parameter.

Each sfermion mass at the GUT scale would be an additional parameter
independent of $\tan\beta$, $M_2$ and $\mu$.
However, assuming a common sfermion mass, $m_0$, at the GUT scale,
the sfermion masses run with a mass scale 
according to the 
renormalization group equations. 
Assuming a common gaugino mass and 
a common sfermion mass at the GUT scale within the framework 
of the MSSM leads to a model called 
constrained MSSM (CMSSM).  
The interpretation of the results in this publication is based 
on the CMSSM, although some results are valid in a larger
framework.
In this model the squark, slepton and sneutrino masses at the 
weak scale are approximately  
given by the following formulae\footnote{These formulae cannot 
be directly applied for the masses of the stop, 
sbottom and stau, since these are affected by the large 
corrections due to ${\mathrm \tilde{f}_L}$-${\mathrm \tilde{f}_R}$ mixing.
More accurate formulae for the sfermion masses  
are provided in \cite{sfermion2}.
} \cite{sfermion1}:
\begin{eqnarray}
m_{\su_R}^2 & = & m_0^2 + 5.87~m_{1/2}^2 
-0.16~m_{\mathrm{Z}}^2 |\cos 2\beta | \\
m_{\su_L}^2 & = & m_0^2 + 6.28~m_{1/2}^2 
-0.35~m_{\mathrm{Z}}^2 |\cos 2\beta | \\
m_{\sd_R}^2 & = & m_0^2 + 5.82~m_{1/2}^2 
+0.08~m_{\mathrm{Z}}^2 |\cos 2\beta | \\
m_{\sd_L}^2 & = & m_0^2 + 6.28~m_{1/2}^2 
+0.42~m_{\mathrm{Z}}^2 |\cos 2\beta | \\
m_{\sell_R}^2 & = & m_0^2 + 0.15~m_{1/2}^2 
+0.23~m_{\mathrm{Z}}^2 |\cos 2\beta | \\
m_{\sell_L}^2 & = & m_0^2 + 0.52~m_{1/2}^2 
+0.27~m_{\mathrm{Z}}^2 |\cos 2\beta | \\
m_{\snu_L}^2 & = & m_0^2 + 0.52~m_{1/2}^2 
-0.50~m_{\mathrm{Z}}^2 |\cos 2\beta |,
\end{eqnarray}
where the last term in each expression is zero for $\tan \beta =1$. 
Limits in the ($M_2$,$\mu$) parameter space, obtained
from the chargino and neutralino searches, are presented for
$\tan \beta = 1.5$ and 35.
These two values of $\tan \beta$  are theoretically interesting since 
at these values the model is consistent with both the measured value of the 
top mass and the mass ratio of the $\tau$ and the bottom quark.
However, $\tan\beta$ could be as small
as 1.0~\cite{tanbone}.
The phenomenology of chargino and neutralino production and decay
changes drastically when $\tan \beta$ approaches 1.0.
Therefore the case of $\tan \beta = 1.0$ is also studied.

For $\tan \beta$ close to 1.0, the ordinary analysis based on large 
missing momentum is insensitive in the region 
$M_2 \approx \mu \approx 0$ of the ($M_2$,$\mu$) 
plane. In this region the two chargino masses are close to 
the mass of the W boson 
($m_{\ch_1} \approx m_{\ch_2} \approx m_{\mathrm W}$), 
the two lightest neutralinos are almost massless\footnote{
Small gluino masses have not been considered here.}
($m_{\nt_1} \approx m_{\nt_2} \approx 0$), 
and the other two neutralino masses are close to the mass of the
Z boson
($m_{\nt_3} \approx m_{\nt_4} \approx m_{\mathrm Z}$).
In this region one of the two lightest neutralinos is 
an almost pure photino and the other one is an almost 
pure higgsino, hence $\nt_1 \nt_2$ production 
in $\ee$ collisions is heavily suppressed. 
The heavy neutralinos $\nt_3$ and $\nt_4$ are mixtures of the
zino and the other higgsino.
To be sensitive to the region near $M_2 = \mu =0$,  
the ALEPH collaboration has studied $\nt_2 \nt_3$ or $\nt_2 \nt_4$ production 
with the subsequent decays 
$\nt_{3,4} \ra \Zrv \nt_{2}$ and $\nt_2 \ra \nt_1 \gamma$
at centre-of-mass energies $\sqrt{s} = 130$ and 136 GeV
\cite{ALEPHnt}. At these energies charginos are too heavy
to be produced in the considered parameter region.
At $\sqrt{s}$ well above the W-pair threshold, the chargino 
pair production cross-section is typically much larger than the neutralino
pair ($\nt_2 \nt_3$ or $\nt_2 \nt_4$) production cross-sections 
near $M_2 = \mu =0$.   
At $\sqrt{s} = 172$~GeV,
the sum of the cross-sections for the four chargino pair production 
processes ($\ee \ra \chp_1 \chm_1$, $\chp_2 \chm_2$ and $\chpm_1 \chmp_2$) 
near $M_2 = \mu = 0$ is as large as 6~pb, whereas the $\WW$
production cross-section is about 13~pb.

In the ($M_2$,$\mu$) region considered here, 
the event shapes of the chargino pair events  
($\chp_1 \chm_1$, $\chp_2 \chm_2$ or $\chpm_1 \chmp_2$) are
similar to those of ordinary $\WW$ events, since 
each chargino decays into an on-mass-shell or almost on-mass-shell W 
with an almost massless neutralino having low momentum.
These events tend to have somewhat larger missing energy than the 
ordinary W pair events, since the neutralinos tend to have 
small, but significant, momentum.
A large neutralino momentum in the rest frame of the chargino is favoured
due to the larger phase space available in the 
two-body decay $\chpm_i \ra {\Wrv}^+ \nt_j$.
On the other hand, the W boson tends to stay near its mass-shell.
These two effects determine the momentum spectrum of the neutralinos.
A search for an excess of $\WW$-like events 
with respect to the Standard Model expectation
(mainly pairs of W bosons) is performed.
For a sneutrino mass smaller than about 100~GeV the 
chargino pair production cross-section is
reduced due to the negative interference between 
$s$-channel annihilation into $\gamma$ or Z
and $t$-channel $\snu_{\mathrm e}$ exchange diagrams. 
Hence for a small sneutrino mass the region of low sensitivity  
in the ($M_2$,$\mu$) plane becomes significantly large.
The present analysis has therefore been designed in such a way that 
a relatively large region around $M_2 = \mu = 0$ is covered.  
The charginos ($\chpm_1$ and $\chpm_2$) decay either into 
$\nt_1 {\Wrv}^{\pm}$ or $\nt_2 {\Wrv}^{\pm}$. 
The light $\nt_2$ decays subsequently  
into $\nt_1 \gamma$ through loop diagrams.
If the sneutrino is lighter than the chargino,
the two-body decay
$\chpm \ra \tilde{\nu}_{\ell} \ell^{\pm}$ would dominate, but
the subsequent decay of
$\tilde{\nu}_{\ell} \ra \nu_{\ell} \nt_j$ leads to
the same final state topology as the leptonic decay via
$\nt_j {\Wrv}^{\pm}$.
The SUSYGEN Monte Carlo generator \cite{SUSYGEN} is used to calculate 
these branching fractions.

In this publication, the OPAL detector is described in Section 2. The
various event simulations which have been used are described in
Section 3. 
Analyses of the various possible signal topologies
are discussed in Section 4 and
results and physics interpretations, both model independent and 
based on the CMSSM,
are given in Section 5.
%
\section{The OPAL Detector}
%
The OPAL detector
is described in detail in \cite{OPAL-detector}; it
is a multipurpose apparatus
having nearly complete solid angle coverage\footnote{
A right-handed coordinate system is adopted,
where the $x$-axis points to the centre of the LEP ring,
and positive $z$ is along  the electron beam direction.
The angles $\theta$ and $\phi$ are the polar and azimuthal angles,
respectively.}.
The central tracking system consists of a silicon microvertex
detector, a vertex drift chamber, a jet chamber and $z$-chambers.
In the range $|\cos\theta|<0.73$, 159 points can be measured in the
jet chamber along each track. At least 20 points on a track can be
obtained over 96\% of the full solid angle.
The whole tracking system is located
inside a 0.435~T axial magnetic field. 
A lead-glass electromagnetic (EM) calorimeter providing acceptance within
$|\cos\theta|<0.984$ together with
presamplers and time-of-flight scintillators
is located outside the magnet coil and at the front of both endcaps.
The magnet return yoke is instrumented for hadron calorimetry (HCAL)
giving a polar angle coverage of $|\cos\theta|<0.99$
and is surrounded by external muon chambers.
The forward detectors (FD) and
silicon tungsten calorimeters\footnote{In 1996, tungsten shields 
were installed around the beam pipe
in front of the SW detectors to reduce the amount of synchrotron radiation
seen by the detector. The presence of the shield results in a hole
in the SW acceptance between the polar angles of 28 and 31~mrads.}
(SW) located on both sides of the interaction point measure the luminosity
and complete the geometrical acceptance
down to 24~mrad in polar
angle. The gap between the endcap EM calorimeter and FD
is filled by an additional electromagnetic calorimeter,
called the gamma-catcher (GC).

%
\section{Event Simulation}
%
The DFGT generator \cite{DFGT} 
was used to simulate signal events.
It includes spin correlations and
allows a proper treatment of the W boson
width effect in the chargino decay, in particular when the chargino
decays into quasi-on-mass-shell W bosons as in the case of 
$\tan \beta=1.0$ with $M_2 \approx \mu \approx 0$. 

The results obtained using the DFGT generator were cross-checked 
using the SUSYGEN generator \cite{SUSYGEN}.
Both generators include initial state radiation.
The JETSET~7.4 package~\cite{PYTHIA}
is used for the hadronisation of the quark-antiquark
system in the chargino or neutralino hadronic decays.
It is important to incorporate correctly all the possible branching fractions 
of charginos and neutralinos.   
SUSYGEN is used to calculate 
these branching fractions.

The most important parameters influencing the chargino detection efficiency
are the mass of the lightest chargino, $m_{\chp_1}$,
and the mass difference between the lightest chargino and the lightest
neutralino, $\Delta M_+ \equiv m_{\chp_1} - m_{\nt_1}$.
For the neutralino detection efficiency, $m_{\nt_2}$ and
$\Delta M_0 \equiv m_{\nt_2} - m_{\nt_1}$ are the main parameters that
affect the efficiency.
For $\chp_1 \chm_1$ events, 64 points were generated
in the ($m_{\chp_1}$,$\Delta M_+$) plane, for
$m_{\chp_1}$ between 50~GeV and 85~GeV and $\Delta M_+$ between 3~GeV
and $m_{\chp_1}$, in the wino-higgsino mixed case and in the
pure higgsino case.
The correspondence of these cases to the MSSM parameters
is explained in Section~\ref{interp}.
To study systematic effects due to variations in the matrix element
which lead to
different production and decay angular distributions,
events were generated at 32 additional points in the pure wino case.
At each point 1000 events for the decay
$\chp_1 \ra \nt_1 {\mathrm W}^{*+} $
were generated.
For $\nt_1 \nt_2$ events, 62 points were generated
in the ($m_{\nt_1}$,$m_{\nt_2}$) plane, for
$(m_{\nt_1}+m_{\nt_2})$ between 100~GeV and 170~GeV and $m_{\nt_1}$
between 10~GeV and $(m_{\nt_2}-3.0$~GeV).
At each point, 1000 events for the decay
$\nt_2 \ra \nt_1 {\mathrm Z}^{*}$ with ${\mathrm Z}^{*} \ra \ellell$ or
$\qq$ were generated.

The sources of background to the chargino and neutralino signals
include two-photon, lepton-pair, multihadronic and four-fermion
processes.
Two-photon processes are the most important background
for the case of small $\Delta M_+$ where
the visible energy and momentum transverse to
the beam direction for the signal and the two-photon events are comparable.
The Monte Carlo generators
PYTHIA~\cite{PYTHIA}, PHOJET~\cite{PHOJET} and HERWIG~\cite{HERWIG}
were used for simulating hadronic events from two-photon
processes.
Other four-fermion processes, excluding \linebreak
$\ee\ee$,
were simulated using the grc4f~\cite{grc4f}
generator, which takes into account all interfering four-fermion
diagrams. 
The dominant contributions are
${\mathrm W}^+ {\mathrm W}^-$, $\Zgv$ or $\Z^*\Z$ events that have
topologies very similar to that of the signal.
The Vermaseren~\cite{Vermaseren} program was used to
simulate $\ee\ee$, as well as additional samples of
$\ee \mumu$ and $\ee \tautau$ processes.
The EXCALIBUR~\cite{excalibur} program was used as a cross-check.
Lepton pairs were generated using
the KORALZ~\cite{KORALZ} generator for
$\tau^+ \tau^- (\gamma)$ and $\mumu (\gamma)$ events
and the BHWIDE~\cite{BHWIDE} program
for $\ee \ra \ee (\gamma) $ events.
Multihadronic, $\qq (\gamma)$, events were simulated using
PYTHIA. 

The simulated background events were all generated 
at $\sqrt{s} = 171.0$~GeV.
In evaluating the number of background events,
the cross-sections were calculated at the actual
centre-of-mass energies (170.3~GeV and 172.3~GeV) and were weighted by
the corresponding collected luminosities.

Generated signal and background events were processed
through the full simulation of the OPAL detector~\cite{GOPAL}
and the same event analysis chain was applied to the simulated events
as to the data.
%
\section{Analysis}
\label{analysis}

The analysis is performed on data
collected during the 1996 autumn run of LEP
at centre-of-mass energies of $\roots = $170.3 and 172.3~GeV.
Data are used from runs in which all the subdetectors
relevant to this analysis were fully operational, corresponding
to an integrated luminosity of $\lumi$~pb$^{-1}$ (1.0~pb$^{-1}$ at
170.3~GeV and 9.3~pb$^{-1}$ at 172.3~GeV).
The luminosity is measured using small angle Bhabha scattering 
events detected in
the silicon tungsten calorimeter.

To select good charged tracks and clusters in the calorimeters,
quality requirements identical to those in Ref.~{\cite{LEP15-opal} are
applied.

Calculations of experimental variables are
performed using
the four-momenta of tracks and of EM or HCAL clusters not
associated with charged tracks\footnote{
The masses of all charged particles are set to
the charged pion mass, and the invariant masses of the calorimeter
energy clusters
are assumed to be zero.}.
Calorimeter clusters associated with charged tracks
are also included after
the expected calorimeter energy for the associated
charged track momenta is
subtracted
from the cluster energy to reduce double counting.
If the energy of a cluster is smaller
than the expected energy for the associated tracks,
the cluster energy is not used.

Jets are formed 
from charged tracks and calorimeter clusters
using the Durham
algorithm~\cite{DURHAM} with
a jet resolution parameter of $y_{\mathrm cut} = 0.005$,
unless otherwise specified.

To select well measured events 
the following preselection criteria are applied:
\begin{itemize}
\item[(P1)]
  The number of charged tracks 
  satisfying the quality criteria is required to be at least two.
  Furthermore, 
  the ratio of the number of tracks satisfying the quality criteria 
  to the total number of reconstructed tracks is required 
  to be larger than 0.2.
\item[(P2)]
  The event transverse momentum relative to the beam direction
  is required to be larger
  than 1.8~GeV.
\item[(P3)]
  The total energy deposited
  in each side of the silicon tungsten calorimeter, 
  forward calorimeter and the gamma-catcher
  has to be smaller than 2~GeV, 2~GeV 
  and 5~GeV, respectively.
\item[(P4)]
  The visible invariant mass of the event has to exceed 3 GeV.
\item[(P5)]
  The maximum EM cluster energy 
  and the maximum charged track momentum both have to be smaller than 
  130\% of the beam energy.
\end{itemize}
After these preselection cuts 6310 data events 
are selected for further analysis
compared with an expectation from the simulation of the relevant 
backgrounds of
5535~events. The difference between observed and expected events is
attributed to the incomplete modelling of low mass two-photon
processes by the available generators. 
As described later, reducing the two-photon contribution in
the course of the analysis yields
satisfactory agreement between data and background simulation.

\subsection{Detection of charginos}

For the chargino search, the event sample is divided into three 
categories, motivated by the topologies expected to result from chargino
events. Separate analyses are applied to the preselected events in 
each category to obtain optimal performance:
\begin{itemize}
\item[(A)] $N_{\mathrm ch} > 4$ and no isolated lepton observed,
where $N_{\mathrm ch}$ is the number of reconstructed charged tracks,
\item[(B)] $N_{\mathrm ch} > 4$ and at least one isolated
lepton observed, and
\item[(C)] $N_{\mathrm ch} \leq 4$.
\end{itemize}

For the preselected events 1448 are classified as (A), 1296 as (B) and
3566 as (C).

These analyses are similar to those used for the analysis of 161~GeV
data \cite{LEP16-opal} but their robustness to the larger $\WW$
background has been significantly improved.

Isolated leptons
are identified in the following way.
Electrons are selected if they satisfied either the artificial neural
network electron identification described 
in~\cite{e1} or the one used for the OPAL $R_{\mathrm b}$ analysis~\cite{e2}. 
Muons are selected if they satisfied one of three muon
identification methods: the first method is based on the best
matching track to the muon chamber
track segment~\cite{mu1}, the second on the            
use of the hadron calorimeter as described in~\cite{mu2} and the third
one is applying
the cuts used in the OPAL Z line shape analysis~\cite{mu3}. 
The momentum of the electron or muon candidate
is required to be larger than 2~GeV.
A reconstructed jet is identified as a tau decay if
there are only one or three charged tracks in the jet,
the momentum sum of the charged tracks is larger than 2~GeV,
the invariant mass of the charged particles in the jet is
smaller than 1.5~GeV and
the invariant mass of the jet
is smaller than the mass of the tau~\cite{PDG}.
The lepton is defined to be isolated if the
energy within a cone of half angle $20\degree$ around
the electron, muon or tau candidate
is less than 2~GeV.

The fractions of $\chp_1 \chm_1$ events falling into
categories (A), (B) and (C) for
various mass combinations of ($m_{\nt_1}$, $m_{\ch_1}$)
are
given in Table \ref{tab:catch}.

\begin{table}[b]
\centering
\begin{tabular}{|l||c||c|c|c|c|c|c|c|}
\hline
$\chp_1 \chm_1$ & $m_{\chp_1}$
     & 50  & 60  & 65  & 70  & 75  & 80  & 85  \\
  &  & GeV & GeV & GeV & GeV & GeV & GeV & GeV \\
\hline
\hline
$\Delta M_{+}$  &Category&       &       &       &       &       &     &  \\
3.0~GeV  & (A) &    3&    3&    3&    3&    5&    5&    7 \\
         & (B) &   10&   11&   10&   10&    9&    8&    7 \\
         & (C) &   87&   86&   88&   87&   87&   87&   86 \\
\hline
 5.0~GeV & (A) &   12&    9&    8&    7&    7&    8&    7 \\
         & (B) &   32&   32&   35&   37&   37&   34&   38 \\
         & (C) &   56&   59&   57&   56&   56&   58&   55 \\
\hline
10.0~GeV & (A) &   28&   20&   20&   17&   15&   13&   12 \\
         & (B) &   45&   50&   55&   53&   55&   55&   61 \\
         & (C) &   27&   30&   25&   30&   30&   32&   27 \\
\hline
20.0~GeV & (A) &   38&   36&   36&   34&   33&   30&   32 \\
         & (B) &   46&   47&   50&   49&   50&   53&   50 \\
         & (C) &   16&   17&   14&   17&   17&   17&   18 \\
\hline
$m_{\chp_1}/2$& (A) &   42&    45&   44&   45&   48&   50& 49 \\
              & (B) &   44&    43&   41&   43&   39&   39& 40 \\
              & (C) &   15&    12&   16&   13&   13&   10& 11 \\
\hline
$m_{\chp_1}$& (A) &   45&   44&   50&   51&   50&   53& 54 \\
$-$20~GeV&    (B) &   40&   43&   40&   36&   38&   35& 35 \\
         &    (C) &   15&   13&   10&   13&   12&   12& 11 \\
\hline
$m_{\chp_1}$& (A) &   48&   51&   50&   49&   54&   54& 59 \\
$-$10~GeV&    (B) &   39&   37&   36&   38&   36&   34& 32 \\
         &    (C) &   13&   11&   13&   13&   11&   11&  9 \\
\hline
$m_{\chp_1}$& (A) &   53&   51&   54&   53&   53&   56& 57 \\
            & (B) &   36&   38&   34&   36&   37&   34& 33 \\
            & (C) &   11&   11&   11&   11&   10&   10& 10 \\
\hline
\end{tabular}
\caption[]{
   \protect{\parbox[t]{12cm}{
The percentages of the simulated $\chp_1 \chm_1$ event samples
falling into each of the three categories
for $\chpm_1 \ra \nt_1\Wvpm$ decay,
where $\Delta M_{+}=m_{\chp_1}-m_{\nt_1}$. These percentages have been
evaluated using 1000 events for each mass combination.
}  }}
\label{tab:catch}
\end{table}

To complete the chargino search in a particular region of SUSY parameter space
($M_2 \approx \mu \approx 0$ for $\tan \beta=1$)
which is not accessible by analyses (A), (B) and (C),
two additional analyses described in Section~\ref{sec:WWevents} 
are performed.

\subsubsection{Analysis A (\boldmath$N_{\mathrm ch} > 4$
without isolated leptons)}
For reasonably large values of $\Delta M_+$,
if both $\chp_1$ and $\chm_1$ decay
hadronically,
signal events tend to fall into category (A).
As listed in Table~\ref{tab:catch},
the fraction of $\chp_1 \chm_1$ events falling into
category (A) is 30--59\% if 
$\Delta M_+$ is 20~GeV or greater.
The fraction drops  to less than or equal to 12\%
for $\Delta M_+ \leq 5$~GeV since the
average charged track multiplicity of these events is small.

 For an event to be considered as a candidate it has to satisfy the
 following criteria:

\begin{itemize}

\item[(A1)] To ensure that the events are well contained, a requirement is
  imposed on the direction of the missing momentum 
  ($\cosmiss <0.9$). Furthermore, the ratio of the
  measured energy in the forward
  cones (defined by $|\cos \theta | > 0.9$) 
  to the total measured energy should be smaller than 15\%.

\item[(A2)]  The background from multihadronic two-photon processes is strongly
  reduced by imposing a cut on the total transverse momentum of the
  event including ($P_t^{\mathrm{HCAL}}>6$~GeV) and not including
  ($P_t>5$~GeV) the information from the hadron
  calorimeter. Although most of the events from two-photon processes
  are rejected by the $P_t$ cut, the $P_t^{\mathrm{HCAL}}$ cut is
  applied to reject two-photon events with an occasional high
  transverse momentum neutral hadron. In addition, the longitudinal
  component of the missing momentum is required to be smaller than
  35~GeV, which reduces the contribution from radiative $\mathrm Z$ events.

\item[(A3)] To reject multihadronic annihilation and $\WW$ final
  states, further requirements are imposed on the acoplanarity and
  measured mass of the event. The event is divided into two jets using
  the Durham jet algorithm. The acoplanarity angle, $\phiacop$,
  is defined as the
  complement of the azimuthal opening angle of the two jets, each of
  which should be well contained ($|\cos \theta |< 0.95$).
  Figure~\ref{fig:cataacop} shows
  the acoplanarity angle distribution for the data and the
  various background components, 
  and for various signals. The acoplanarity angle should exceed
  $15 \degree$. 
  Furthermore, the total observed mass of the event should be smaller than
  100 GeV.

\item[(A4)] Remaining $\WW$ events where one W decays semileptonically are
  rejected by using a different lepton algorithm which has less
  stringent isolation criteria [3]. Events containing a lepton of this
  type, with a hadronic system recoiling against it with a reconstructed
  mass between 55 to 85~GeV are rejected.
\end{itemize}

 Events are then classified according to the number of hadronic jets.
 Figure~\ref{fig:catamvis}(a) shows the number of hadronic jets versus 
 reconstructed invariant mass for
 $\ellell \qq$, $\nulqq$ and
 $\nunu{\mathrm q} \bar{\mathrm q}$ final state 
 background events that are the dominant background processes at this 
 stage. Most of the events fall into the
 two- and three-jet topologies. Figures~\ref{fig:catamvis}(b) and (c) 
 show the same distributions for
 the expected signal events in the case of small
 $\Delta M_{+}$
 and large $\Delta M_{+}$, respectively. It can
 be seen that while the low $\Delta M_{+}$ case is dominated by low
 masses and two or
 three jets, the large $\Delta M_{+}$ case is dominated by four-jet events
 with large invariant mass.
 To preserve high detection efficiency for both large and small 
 $\Delta M_{+}$, the analysis is divided
 into two parts.

 In the first part, two- and three-jet events are selected, and the
 following requirements are imposed:

\begin{itemize}
\item[(A5a)] Highly boosted events satisfying $\Mvis / \Evis<0.5$ are rejected.
     This requirement reduces the contribution from the background
     due to the reaction 
     $\ee \rightarrow \mathrm{Z} \gamma^* \rightarrow \nunu \qq $.

\item[(A6a)] The energy of the most energetic jet should not exceed 35~GeV
     and the observed mass should not exceed 56~GeV (63~GeV) for two-jet
     (three-jet) events. These requirements strongly reduce the
     remaining background from $\WW$ and $\Wenu$ production, since 
     their observed
     mass peaks at 80~GeV (see Fig.~\ref{fig:catamvis}).

\end{itemize}

 Events with four or more jets are selected in the second part.
 Events with a clear four-jet signature are selected by requiring
 the fourth-most energetic jet to have an energy exceeding 8~GeV and 
 that each jet contains at least one
 charged particle. These events are rejected if they change
 from being three-jet to four-jet events at $y_{\mathrm cut} = y_{34}$
 smaller than 0.01,
 or if they change from being four-jet to five-jet events at a
 $y_{45}$ value larger than 0.0015.

\begin{itemize}

\item[(A5b)] Events are required to have either a clear four-jet
  signature (as defined above) or no jet with energy
  exceeding 20~GeV.

\end{itemize}

No events in the data survive the cuts described above. This is consistent
with the expected background from Standard Model processes of 0.24~events. 
The numbers of events remaining after each cut are listed
in Table~\ref{tab:nevA}. After cut (A1), which
rejects multihadronic $\gamma \gamma$ events, there is good
agreement between the data and
the Standard Model predictions.

For events falling into category (A) the
efficiencies for $\chp_1 \chm_1$ events are listed
in Table \ref{tab:effAchWv}
for the $\nt_1\Wvpm$ decay of the $\chpm_1$. The numbers for 
$\Delta M_{+}\leq 5$~GeV suffer from large statistical
fluctuations but these do not matter for the final results
as very few events fall into category (A) when $\Delta M_{+}$ is small.

\begin{table}[t]
\centering
\begin{tabular}{|l||r||r||r|r|r|r||r|r|}
\hline
   & data & total & $\qq (\gamma)$ & $\ellell (\gamma)$&
`$\gamma \gamma$' & 4-f & $\chp_1 \chm_1$ & $\chp_1 \chm_1$ \\
   &      & bkg.  &                &                   &
                  &     &                 &                 \\
\hline
$m_{\ch_1}$ (GeV)&       &      &         &     &       &
        & 80 & 80 \\
$m_{\nt_1}$ (GeV)&       &      &         &     &      &
        & 60 & 20 \\
\cline{1-1}
 cut &     &      &                  &           &       &
    &  &     \\
\hline
\hline
no cuts  &     -- &    -- &     -- &      -- &     -- &     --  & 1000 & 1000 \\
\hline
Presel.+(A) & 1448 &  1282.  &  777.  &   13.7 &   408.   &  83.0 &   280 &   491 \\  
\hline
Cut (A1) &  423  &  407. &   267.   &   2.54  &   80.9 &    56.4 &   224 &   375 \\  
\hline
Cut (A2) & 193 &   196.  &  156.   &   1.73   &    0.72   &  37.7  &  217  &  339 \\  
\hline
Cut (A3) &   8  &    4.50     &   0.15   &    0.01    &   0.22 &     4.11 &   197 &   216 \\  
\hline
Cut (A4) &   7   &   3.84    &   0.15     &  0.01   &    0.22   &   3.45  &  197  &  192 \\  
\hline
  & \multicolumn{8}{|c|}{Events with two or three jets} \\
\hline
2 or 3 jets &   5  &    2.94  &     0.08   &    0.01  &     0.08   &   2.76 &  121 & 107   \\

\hline
Cut (A5a) &   4  &    2.75   &    0.08   &    0.007   &    0.08  &    2.57 &   116 &   103 \\  
\hline
Cut (A6a) &   0  &    0.18   &    0.00    &   0.002   &    0.08  &    0.09  &   50   &  36 \\  
\hline
  & \multicolumn{8}{|c|}{Events with at least four jets} \\
\hline
$\geq$ 4 jets &    1   &    0.65    &   0.07  &     0.00  &  0.00 & 0.58 &  75  &   85  \\ 
\hline
Cut (A5b) &   0  &    0.06 &      0.006   &    0.00   &    0.00   &    0.06  &  65  & 51 \\
\hline
  & \multicolumn{8}{|c|}{Final Counts} \\
\hline
Net &    0   &   0.24   &   0.006  &     0.002  &     0.08  &0.15 &   115   &  87 \\
\hline
\end{tabular}
\caption[]{
   \protect{\parbox[t]{12cm}{
The remaining numbers of events
after each cut
for various background processes normalized to $\lumi$~pb$^{-1}$
are compared with data for category (A).
Numbers for
two simulated event samples of $\chp_1 \chm_1$ with
$\chp_1 \ra \nt_1 {\mathrm{W}}^*$ are
also given.
 }} 
}
\label{tab:nevA}
\end{table}

\begin{table}[b]
\centering
\begin{tabular}{|l||r|r|r|r|r|r|r|r|}
\hline
$m_{\chp_1}$ (GeV) & 50 & 55 & 60 & 65 & 70 & 75 & 80 & 85 \\
\hline
\hline
$\Delta M_{+}$       &       &       &       &    &  &   &  &       \\
\hline
 3.0~GeV
 &   7 &  12 &   6 &  12  &  0  &  2  &  0 &  0 \\
\hline
 5.0~GeV
 &  16 &  29 &  27 &  28 &  15 &  23  &  7  & 28 \\
\hline
 10.0~GeV
 &  47 &  50 &  47 &  48 &  51 &  45 &  53 &  51 \\
\hline
 20.0~GeV
 &  28 &  33 &  35 &  43 &   49 &  54 &  58 &  58 \\
\hline
 $m_{\chp_1}/2$
 &  19 &   19 &  16 &  24 &  23 &  25 &  26 &  22 \\
\hline
 $m_{\chp_1}-20$~GeV
 &  15 &  14 &  13 &  13 &  15 &  14 &  16 &  13 \\
\hline
 $m_{\chp_1}-10$~GeV
 &  11 &  10 &  11  &  9  &  8 &   9 &   6  &  6 \\
\hline
 $m_{\chp_1}$
 &  8 &   5 &   6 &   5 &   5  &  5 &   3 &   0 \\
\hline
\end{tabular}
\caption[]{
   \protect{\parbox[t]{12cm}{
The detection efficiencies for events falling into
category (A), in percent, for $\chp_1 \chm_1$
with two $\chpm_1 \ra \nt_1 \Wvpm$ decays normalized to the number of
events without isolated leptons and with $N_{\mathrm{ch}}>4$.
 }} 
}
\label{tab:effAchWv}
\end{table}

\subsubsection{Analysis B (\boldmath$N_{\mathrm ch} > 4$ with isolated leptons)}

In $\chp_1 \chm_1$ events in which one of the $\chpm_1$ decays
leptonically, the events tend to fall into category (B).
The fraction of $\chp_1 \chm_1$ events falling into
category (B) is 32--61\% if $\Delta M_{+}$ is at least 5~GeV.

To reduce the background from $\ee \ra \Zg$ events and 
events from two-photon processes
the following cuts are applied:
\begin{itemize}
\item[(B1)]
The visible energy in the region $|\cos \theta|>0.9$
should be less than 20\% of the total visible energy.
The polar angle of the missing momentum direction, $\thmiss$,
is required to satisfy $\cosmiss < 0.9$.
The polar angle of the thrust axis direction, $\thethr$,
is also required to satisfy $\acosthr < 0.9$.
\item[(B2)]
$P_t$ and $P_t^{\mathrm{HCAL}}$ should be
greater than 5~GeV and 6~GeV, respectively.
The distribution of $P_t^{\mathrm{HCAL}}$
is shown in Fig.~\ref{figb1} after cut (B1).  
\item[(B3)]
The acoplanarity angle as defined in cut (A3)
is required to be greater than 20$\degree$. 
\end{itemize}

\noindent
To reduce the contribution of events from W boson 
production the following three cuts are applied.
\begin{itemize}
\item[(B4)]
The invariant mass of the event excluding the highest energy lepton,
$M_{\mathrm rest}$, should be smaller than 65~GeV.
\clearpage
\item[(B5)]
Events are rejected if the reconstructed highest lepton energy,
$E^{\mathrm max}_\ell$, exceeds 30~GeV.
The scatter plots of $M_{\mathrm rest}$ vs.\ $E^{\mathrm max}_\ell$ 
after cut (B3) are shown in Fig.~\ref{figb3}.
\item[(B6)]
The invariant mass of the event
is required to be smaller than 75~GeV
to reduce contributions from $\WW$ and $\Wenu$ processes.
This cut suppresses W-pair events which decay to 
$\ell \nu \mathrm{q \bar{q'} g}$, in which the charged lepton escapes
down the beampipe and a hadronic jet from the $\mathrm{ q \bar{q'} g}$
system is misidentified as a tau lepton; these are not removed by cuts
(B4) or (B5). 
\end{itemize}

The numbers of events remaining after each cut are shown in 
Table~\ref{tab:nevB} for the OPAL data, the Monte Carlo simulations of
the various background sources and
for three samples of simulated $\chp_1 \chm_1$ events.
There are no events passing all selection cuts, while the expected
background is 0.29~events.
For events falling into category (B)
the efficiencies for $\chp_1 \chm_1$ events are listed
in Table \ref{tab:effBchWv}
for the $\nt_1\Wvpm$ decay of the $\chpm_1$.

\begin{table}[b]
\centering
\begin{tabular}{|l||r||r||r|r|r|r||r|r|r|}
\hline
   & data & total &
$\qq (\gamma)$ & $\ellell (\gamma)$ &
\multicolumn{1}{c|}{`$\gamma \gamma$'} &
\multicolumn{1}{c||}{4-f} &
\multicolumn{3}{c|}{$\chp_1 \chm_1$}  \\
\hline
$m_{\ch_1}$ (GeV)&   &     &         &                    &            &
        & \multicolumn{1}{c|}{80} &
          \multicolumn{1}{c|}{80} &
          \multicolumn{1}{c|}{80}  \\
$m_{\nt_1}$ (GeV)&   &    &         &                    &            &
        & \multicolumn{1}{c|}{70} &
          \multicolumn{1}{c|}{40} &
          \multicolumn{1}{c|}{20}  \\
\cline{1-1}
 cut & & & & & & & & & \\
\hline
\hline
 no cuts
&      &       &       &       &       &       &1000 &1000 &1000 \\
\hline
 Presel.+(B) 
& 1296 &  1027. &  18.9 &  24.9 &   939. &  44.0 & 518 & 378 & 329 \\
\hline
 Cut (B1)
&  213 &   196. &  3.81 &  5.86 &   155. &  30.9 & 430 & 296 & 269  \\
\hline
 Cut (B2)
&   36 &  37.3 &  2.48 &  4.27 &  0.53 &  30.0 & 323 & 286 & 264  \\
\hline
 Cut (B3)
&   14 &  17.4 & 0.07 & 0.12 & 0.00 &  17.2 & 308 & 241 & 223  \\
\hline
 Cut (B4)
&    1 &  3.77 & 0.04 & 0.08 & 0.00 &  3.65 & 308 & 220 & 168  \\
\hline
 Cut (B5)
&    0 & 0.45 & 0.01 & 0.04 & 0.00 & 0.40 & 308 & 214 & 163  \\
\hline
 Cut (B6)
&    0 & 0.29 & 0.01 & 0.03 & 0.00 & 0.25 & 308 & 208 & 154  \\
\hline
\end{tabular}
\caption[]{
  \protect{\parbox[t]{12cm}{
The remaining numbers of events for data and
for various background processes normalized to $\lumi$~pb$^{-1}$
are compared after each cut in category (B).
Numbers for
three simulated event samples of $\chp_1 \chm_1$ with
$\chp_1 \ra \nt_1 {\mathrm{W}}^*$
are also given.
}} }
\label{tab:nevB}
\end{table}

\begin{table}[ht]
\centering
\begin{tabular}{|l||r|r|r|r|r|r|r|r|}
\hline
~~~~~~~$m_{\chp_1}$ (GeV) & 50 & 55 & 60 & 65 & 70 & 75 & 80 & 85 \\
\hline
\hline
$\Delta M_{+}$       &     &     &     &     &     &     &     &     \\
\hline
 3.0~GeV             &   2 &   6 &   3 &   2 &   2 &   2 &   0 &  0  \\
\hline
 5.0~GeV             &  22 &  27 &  25 &  21 &  18 &  20 &  19 & 22  \\
\hline
 10~GeV              &  43 &  47 &  54 &  54 &  53 &  55 &  56 & 58  \\
\hline
 20~GeV              &  49 &  55 &  58 &  61 &  56 &  61 &  69 & 63  \\
\hline
 $m_{\chp_1}/2$      &  51 &  50 &  48 &  53 &  52 &  52 &  53 & 60  \\
\hline
 $m_{\chp_1}-20$~GeV &  42 &  41 &  42 &  42 &  45 &  41 &  44 & 41  \\
\hline
 $m_{\chp_1}-10$~GeV &  34 &  41 &  35 &  34 &  35 &  32 &  27 & 23  \\
\hline
 $m_{\chp_1}$        &  33 &  31 &  30 &  27 &  19 &  19 &  11 &  2  \\
\hline
\end{tabular}
\caption[]{
  \protect{\parbox[t]{12cm}{
The detection efficiencies in percent for $\chp_1 \chm_1$
with two $\chpm_1 \ra \nt_1 \Wvpm$ decays normalized to the number of
events in category (B).
}} }
\label{tab:effBchWv}
\end{table}

%

\subsubsection{Analysis C (\boldmath$N_{\mathrm ch} \leq 4$)}
\label{catccuts}
Events in which both charginos decay hadronically, but with small
$\Delta M_+$, and events in which both charginos decay leptonically tend
to fall into category (C).
The fraction of $\chp_1 \chm_1$ events falling into
category (C) is 9--32\% for $\Delta M_+ \geq 10$~GeV
and $\geq 55$\% for $\Delta M_+ \leq 5$~GeV.
This analysis will also be used in the neutralino search as described
later.

Events are forced into two jets using the Durham jet algorithm~\cite{DURHAM} 
and are required to satisfy the following cuts on the jet and
event variables:
\begin{itemize}
\item[(C1)]
One of the jets must have a transverse momentum greater than 
1.5~GeV and the other must have a transverse momentum greater than 1.0~GeV.
\item[(C2)]
A cut is applied to $P_t$ (shown in Fig.~\ref{catc1}) 
at a value that depends on the acoplanarity angle. The separation by
acoplanarity angle is needed to reduce efficiently 
the background from tau pairs
that occurs at low acoplanarity angles. 
To reject these events, which tend to be back-to-back, a cut
on the component of $P_t$ transverse to the thrust axis, $a_t$,
is applied if the acoplanarity angle is small.
Even when $P_t$ is large, $a_t$ is relatively small for
tau pairs as compared to the signals.
For events with $\phiacop<$50$\degree$ it is required that 
$\axicos$ be less than 0.95
(where $\theta_a^{\mathrm miss}= \tan^{-1}(a_t/P_z)$ and $P_z$ is the
longitudinal component of the missing momentum),
$P_t /E_{\mathrm beam}$ be greater than 0.035, and that $a_t/E_{\mathrm beam}$
exceed 0.025.
For events with $\phiacop>$50$\degree$ it is required that
$|\cos\theta_{\mathrm miss}|$ be less than 0.90,
and that $P_t$$/E_{\mathrm beam}$ exceed 0.050.
\item[(C3)]
To reduce the $\ee \mumu$ background, events
are rejected if there is evidence in the muon chambers, hadron calorimeter
strips or central detector
of a muon escaping in the very forward region,
back to back (within 1 rad) with the direction of the
momentum sum of the dijet system.
Also, events are rejected if there is a relatively large fraction of
hadronic energy ($E_{\mathrm{HCAL}} > 0.05 E_{\mathrm{tracks}}$)
in the event,
where $E_{\mathrm{tracks}}$ is the sum of the energy of the good tracks and
$E_{\mathrm{HCAL}}$ is the sum of the energy 
in the hadron calorimeter clusters.
Lastly the events must be electrically neutral and neither jet
may have a charge of magnitude exceeding 1.
\item[(C4)]
The two-photon background is further reduced by rejecting events if
either of the
jets has $|\cos\theta|>0.75$.
\item[(C5)]
The acoplanarity angle 
is required to be greater than 30$\degree$. This removes
much of the $\ell^+\ell^-\gamma$ background.
\item[(C6)]
To remove the W-pair background, events are rejected if one of the jets
has an energy greater than 22~GeV. The distributions of the energy of
the higher energy jet are shown
in Fig.~\ref{catc3} after all other cuts.
\end{itemize}

The total background predicted by the Standard Model is 0.34 events
for $\lumi$~pb$^{-1}$.
The numbers of events passing each cut are given in Table~\ref{tab:alcbkg}
for data, background and two simulated signal samples.
For events falling into category (C)
the efficiencies for $\chp_1 \chm_1$ events are listed
in Table \ref{tab:effCchWv}
for the $\nt_1 \Wvpm$ decay of the $\chpm_1$.
No data events survive the category (C) cuts.

\begin{table}[t]
 \centering
 \begin{tabular}{|l||r||r||r|r|r|r||r|r|}
 \hline
   &  \multicolumn{1}{c||}{data} & total &
$\qq (\gamma)$ & $\ellell (\gamma)$ & `$\gamma \gamma$' &
\multicolumn{1}{c||}{4-f} &
$\chp_1 \chm_1$ & $\nt_1 \nt_2$  \\
     &       & bkg. &
  &  &  &  &  & \\
 \hline
 $m_{\ch_1}$ (GeV)& & & & & & & \multicolumn{1}{c|}{70} & --  \\
 $m_{\nt_1}$ (GeV)& & & & & & & \multicolumn{1}{c|}{65} &
                                                   \multicolumn{1}{c|}{60}  \\
 $m_{\nt_2}$ (GeV) & & & & & & & \multicolumn{1}{c|}{--} &
                                                   \multicolumn{1}{c|}{70}  \\
 \cline{1-1}
 cut &       &      &         &                    &
         &  &  & \\
 \hline
\hline
no cuts & -- & -- &   -- &  -- & -- &    -- & 1000 & 1000 \\
\hline
 Presel.+(C) & 3566 & 3195. & 0.512 & 2004. & 1173. & 17.9 & 397 & 341 \\
 \hline
 Cut (C1) & 2335 & 2204. & 0.332 & 1759. & 429. & 15.8 & 286 & 234 \\
 \hline
 Cut (C2) & 16 & 18.0 & 0.01 & 5.77 & 2.24 & 9.97 & 158 & 175 \\
 \hline
 Cut (C3) & 8 & 10.1 & 0.01 & 3.16 & 0.70 & 6.20 & 106 & 118 \\
 \hline
 Cut (C4) & 4 & 6.65 & 0.00 & 2.27 & 0.26 & 4.12 & 79 & 83 \\
 \hline
 Cut (C5) & 1 & 3.58 & 0.00 & 0.15 & 0.25 & 3.18 & 74 & 83 \\
 \hline
 Cut (C6) & 0 & 0.34 & 0.00 & 0.006 & 0.25 & 0.08 & 74 & 83 \\
 \hline
\end{tabular}
\caption[]{
\protect{\parbox[t]{12cm}{
For category (C), which is used in the searches for both charginos and
neutralinos,
the remaining numbers of events after each cut are compared with 
various background
processes normalized to $\boldmath\lumi$~pb$^{-1}$.
 Numbers for
 simulated event samples of $\chp_1 \chm_1$ with
 $\chp_1 \ra \nt_1 {\mathrm{W}}^*$
 and $\nt_1 \nt_2$
 with $\nt_2 \ra \nt_1 {\mathrm{Z}}^{*}$ are
 also given.
 }}
}
\label{tab:alcbkg}
 \end{table}

\begin{table}[b]
\centering
\begin{tabular}{|l||r|r|r|r|r|r|r|r|}
\hline
~~~~~~~$m_{\chp_1}$~(GeV) & 50 & 55 & 60 & 65 & 70 & 75 & 80 &85\\
\hline
\hline
$\Delta M_{+}$       &       &     &    &       &       &       &      & \\
\hline
 3.0~GeV         &    5&    5&    4&    6&    4&    2&    3&    2\\
\hline
 5.0~GeV         &   11&   13&   11&   11&   13&   11&    9&   13\\
\hline
 10.0~GeV         &   12&   15&   14&   17&   14&   16&   20&   18\\
\hline
 20.0~GeV         &    9&   13&    7&   10&   13&   15&   20&   18\\
\hline
 $m_{\chp_1}/2$        &    7&    8&   11&    6&    7&    9&   13&    8\\
\hline
  $m_{\chp_1}-20$~GeV  &    4&    6&    8&   10&    9&    5&    5&    1\\
\hline
  $m_{\chp_1}-10$~GeV  &    2&    8&    2&    4&    4&    4&    4&    2\\
\hline
  $m_{\chp_1}$         &    2&    6&    9&    3&    3&    3&    2&    3\\
\hline

\end{tabular}
\caption[]{
  \protect{\parbox[t]{12cm}{
The detection efficiencies in percent for $\chp_1 \chm_1$
with $\chpm_1 \ra \nt_1\Wvpm$ decay normalized to the number of
category (C) events.
}}
}
\label{tab:effCchWv}
\end{table}

\clearpage

\subsubsection{Combined efficiencies and backgrounds 
               for ${\bf \chp_1 \chm_1}$}

The overall efficiency for each mass pair combination is obtained
by taking the sum of the efficiencies for categories (A), (B) and
(C) weighted by the
fraction of signal events falling into each category.
Overall efficiencies for $\chp_1 \chm_1$
events are given in
Table~\ref{tab:effchWv}.
As shown in this table, the efficiencies are 29--57\%
if the mass difference between $\ch_1$ and $\nt_1$ is
$\geq 10$~GeV and $\leq m_{\ch_1}/2$.
The efficiency at an arbitrary point of $m_{\chp_1}$
and $m_{\nt_1}$ is obtained by interpolation using a polynomial fit to
the efficiencies determined from the Monte Carlo.
The total background expected for this search is the sum of the background
contributions from each category. The total background expected for
$\lumi$~pb$^{-1}$ is 0.87~events, consistent with no events being observed in
the data after all cuts.

\begin{table}[ht]
\centering
\begin{tabular}{|l||r|r|r|r|r|r|r|r|}
\hline
~~~~~~~$m_{\chp_1}$ (GeV) & 50 & 55 & 60 & 65 & 70 & 75 & 80 & 85 \\
\hline
\hline
$\Delta M_{+}$       &     &     &     &     &     &     &     &     \\
\hline
 3.0~GeV             &   5 &   5 &   4 &   6 &   4 &   2 &   2 &  1  \\
\hline
 5.0~GeV             &  15 &  19 &  17 &  16 &  15 &  15 &  12 & 17  \\
\hline
 10~GeV              &  36 &  39 &  41 &  43 &  41 &  42 &  44 & 46  \\
\hline
 20~GeV              &  35 &  40 &  41 &  47 &  46 &  51 &  57 & 53  \\
\hline
 $m_{\chp_1}/2$      &  31 &  32 &  29 &  33 &  33 &  34 &  35 & 35  \\
\hline
 $m_{\chp_1}-20$~GeV &  24 &  24 &  25 &  24 &  25 &  23 &  25 & 22  \\
\hline
 $m_{\chp_1}-10$~GeV &  19 &  23 &  19 &  17 &  18 &  17 &  13 & 11  \\
\hline
 $m_{\chp_1}$        &  16 &  15 &  15 &  13 &  10 &  10 &   6 &  1  \\
\hline
\end{tabular}
\caption[]{
  \protect{\parbox[t]{12cm}{
The detection efficiencies in percent combined for the three categories
for $\chp_1 \chm_1$ followed by the decay $\chpm_1 \ra \nt_1 \Wvpm$.
}}
}
\label{tab:effchWv}
\end{table}

\subsection{Detection of neutralinos}

The search for neutralinos is performed by dividing the event sample
into two categories:

\begin{itemize}
\item[(C)] $N_{\mathrm ch} \leq 4$.
\item[(D)] $N_{\mathrm ch} > 4$.
\end{itemize}

Events falling into category (D) have a monojet topology and the
cuts provide better performance for $\nt_1\nt_2$
detection than would have been obtained using the cuts of
categories (A) and (B).
For events with $N_{\mathrm ch} \le 4$ the category (C) cuts, as described in
\ref{catccuts}, are used.

The fractions of
simulated $\nt_1 \nt_2$ events falling into each of
the two categories for $\nt_2 \ra \nt_1\Zrv$ decay are shown
in Table~\ref{tab:catntcd}. The fraction of events falling into
category (C) is 28--42\% for $\Delta M_0 \geq$~20~GeV but
increases to above 85\% when  $\Delta M_0 \leq$~5~GeV.
The efficiencies for $\nt_1 \nt_2$ events for the $\nt_1 \Zv$ decay
of $\nt_2$, normalized to $\nt_1 \nt_2 \ra \nt_1\Zv~(\Zv\ra\qq,\ellell)$,
are listed in Table \ref{tab:effCntZv}.
In category (C) 20\% (absolute fraction) of the events are
invisible due to $\nt_2 \ra \nt_1\Zrv \ra \nt_1 \nunu$ decays.

\begin{table}[ht]
\centering
\begin{tabular}{|l||c||r|r|r|r|r|r|r|r|}
\hline
 $\nt_2 \nt_1$ & $(m_{\nt_2}+m_{\nt_1})$~GeV
& 100 & 110 & 120 & 130 & 140 & 150 & 160 &170  \\
\hline
\hline
$\Delta M_{0}$  &Category &    &    &    &    &    &    &     &    \\
\hline
 3.0~GeV &(C)       &  98 & 98 & 99 & 98 & 99 & 99 & 99 & 99 \\
         &(D)       &   2 &  2 &  1 &  2 &  1 &  1 &  1 &  1 \\
\hline
 5.0~GeV &(C)       &  86 & 85 & 88 & 87 & 87 & 89 & 88 & 89 \\
         &(D)       &  14 & 15 & 12 & 13 & 13 & 11 & 12 & 11 \\
\hline
 10~GeV  &(C)       &  60 & 62 & 59 & 62 & 58 & 60 & 58 & 59 \\
         &(D)       &  40 & 38 & 41 & 38 & 42 & 40 & 42 & 41 \\
\hline
 20~GeV  &(C)       &  41 & 42 & 40 & 39 & 39 & 38 & 40 & 40 \\
         &(D)       &  59 & 58 & 60 & 61 & 61 & 62 & 60 & 60 \\
\hline
 30~GeV  &(C)       &  34 & -- & 34 & -- & 35 & -- & 34 & -- \\
         &(D)       &  66 & -- & 66 & -- & 65 & -- & 66 & -- \\
\hline
50~GeV   &(C)       &  -- & 30 & -- & 33 & -- & 30 & -- & 31 \\
         &(D)       &  -- & 70 & -- & 67 & -- & 70 & -- & 69 \\
\hline
 70~GeV  &(C)       &  30 & -- & 30 & -- & 31 & -- & 31 & -- \\
         &(D)       &  70 & -- & 70 & -- & 69 & -- & 69 & -- \\
\hline
 80~GeV  &(C)       &  30 & 31 & 30 & 30 & 31 & 30 & 30 & 30 \\
         &(D)       &  70 & 69 & 70 & 70 & 69 & 70 & 70 & 70 \\
\hline
 90~GeV  &(C)       &  -- & 28 & -- & 30 & -- & 29 & -- & 30 \\
         &(D)       &  -- & 72 & -- & 70 & -- & 71 & -- & 70 \\
\hline
 110~GeV &(C)       &  -- & -- & 28 & -- & 31 & -- & 29 & -- \\
         &(D)       &  -- & -- & 72 & -- & 69 & -- & 71 & -- \\
\hline
 130~GeV &(C)       &  -- & -- & -- & -- & -- & 30 & -- & 31 \\
         &(D)       &  -- & -- & -- & -- & -- & 70 & -- & 69 \\
\hline
 150~GeV &(C)       &  -- & -- & -- & -- & -- & -- & 30 & -- \\
         &(D)       &  -- & -- & -- & -- & -- & -- & 70 & -- \\
\hline
\end{tabular}
\caption[]{
  \protect{\parbox[t]{12cm}{
The percentages of the simulated $\nt_1 \nt_2$ event samples
falling into each of
the two categories for $\nt_2 \ra \nt_1\Zrv$ decay,
where $\Delta M_{0}=m_{\nt_2}-m_{\nt_1}$.
In category (C) 20\% (absolute fraction) of the events are
invisible events due to $\nt_2 \ra \nt_1\Zrv \ra \nt_1 \nunu$ decays.
}}
}
\label{tab:catntcd}
\end{table}

\begin{table}[ht]
\centering
\begin{tabular}{|l||r|r|r|r|r|r|r|r|}
\hline
\hline
 $(m_{\nt_2}+m_{\nt_1})$~(GeV)
&  100 &110 &120 &130 &140 &150 &160 & 170 \\
\hline
\hline
  $\Delta M_{0}$     &     &     &     &     &     &     &   &  \\
\hline
 3.0~GeV         &    1&    1&    0&    0&    0&    0&    0&    0\\
\hline
 5.0~GeV         &    8&    7&    6&    5&    4&    3&    1&    1\\
\hline
10.0~GeV         &   13&   14&   16&   16&   14&   14&   14&   16\\
\hline
20.0~GeV         &   12&   13&   16&   18&   15&   16&   17&   18\\
\hline
30.0~GeV         &    6&  -- &    9&  -- &   14&  -- &   23&  -- \\
\hline
50.0~GeV         &  -- &    3&  -- &    5&  -- &    8&  -- &    6\\
\hline
70.0~GeV         &    2&  -- &    2&  -- &    2&  -- &    1&  -- \\
\hline
90.0~GeV         &  -- &    0&  -- &    0&  -- &    2&  -- &    0\\
\hline
110.0~GeV         &     &  -- &    2&  -- &    1&  -- &    1&  -- \\
\hline
130.0~GeV         &     &     &     &  -- &  -- &    1&  -- &    0\\
\hline
150.0~GeV         &     &     &     &     &     &  -- &    0&  -- \\
\hline

\end{tabular}
\caption[]{
  \protect{\parbox[t]{12cm}{
The detection efficiencies in percent for $\nt_1 \nt_2$
with $\nt_2 \ra \nt_1\Zv$ decay (with $\Zv \ra \ellell$ or $\Zv \ra \qq$)
normalized to the
number of category (C) events excluding the $\nt_2 \ra \nt_1 \nu\nu$
invisible events.
}}
}
\label{tab:effCntZv}
\end{table}

\subsubsection{Analysis D ($\bf N_{\mathrm ch} > 4$  Neutralino selection)}
In $\nt_1 \nt_{2,3}$ events,
if the $\nt_{2,3}$ decays hadronically,
the events tend to fall into category (D).
Events have to satisfy the following cuts:

To reduce the background from $\ee \ra \Zg$ and two-photon processes
the following cuts are applied.
\begin{itemize}
\item[(D1)]
The visible energy in the region $|\cos \theta|>0.9$
should be less than 15\% of the total visible energy.
The polar angle of the missing momentum direction $\thmiss$
is required to satisfy $\cosmiss < 0.9$\@.
\item[(D2)]
$P_t$ and $P_t^{\mathrm{HCAL}}$ should be
greater than 5~GeV and 6~GeV, respectively.
\item[(D3)]
The acoplanarity angle as defined in cut (A3)
is required to be greater than 15$\degree$. 
Both jets should have a polar angle in the range $|\cos\theta |<0.95$.
Figure~\ref{figdacop} shows the distribution of $\phi_{\mathrm{acop}}$ for the
various background processes and signal samples after cut (D2)\@.
\end{itemize}

\noindent
After these cuts, the remaining background events come 
predominantly from
$\Zgv (\ra \nunu\qq)$, $\WW (\ra \ell \nu\qq^{'})$
and $\Wenu (\ra  \qq ^{'}{\mathrm e} \nu)$\@.
If the invariant mass of the event is smaller than 20~GeV,
the following cut is applied to reduce the contribution of
events from the $\Zgv \ra \nunu\qq$ process:
\begin{itemize}
\item[(D4)]
The ratio of visible mass to visible energy,
$M_{\mathrm vis}$/$E_{\mathrm vis}$, is required to be larger than 0.4\@.
\end{itemize}

\noindent
If the visible mass is greater than 20~GeV,
the following four cuts are applied to reduce
the background from the $\WW$ and $\Wenu$ processes:
\begin{itemize}
\item[(D5.1)]
To keep high efficiency for all lepton flavours, a simple
inclusive leptonic jet identification is used.
Events are forced to be reconstructed into three jets.
The lowest charged multiplicity, $N_{\mathrm min}$, jet is defined as
a `leptonic jet',
and the other two jets as `hadronic jets'.
If there is more than one jet with charged multiplicity equal to
$N_{\mathrm min}$, then
the low multiplicity jet with the largest energy
is defined as the leptonic jet.
The energy of the leptonic jet, ${E_{\mathrm lep}}$,
and the invariant mass of the two hadronic jets, ${M_{\mathrm had}}$,
are required to satisfy
$(M_{\mathrm had} + 2.2 \times E_{\mathrm lep}) \leq $ 110~GeV\@.
Figure~\ref{figdmhad} shows the scatter plot of $M_{\mathrm had}$ versus
$E_{\mathrm lep}$ just before cut (D5.1).

\item[(D5.2)]
If part of the jet escapes undetected 
down the beampipe in a $\WW \ra \tau \nu \qq$
event, the event will have small visible hadronic mass
and survive cut D5.1.
To eliminate such events, the condition
$y_{23} E_{\mathrm vis}^2 < 40\mathrm~GeV^2$ is imposed.

\item[(D5.3)]
Since the topology of the signal events is very similar to that for
$\Wenu$ events, it is very difficult to separate the signal from
background by global kinematical cuts.
Using the decay length method, a loose b-tagging \cite{Higgs161}
is applied when ${M_{\mathrm vis}}$ is larger
than 60~GeV to reduce the contribution of
events from the $\Wenu$ background process
and retain some efficiency for
$\nt_2 \ra \nt_1 {\mathrm b}\bar{\mathrm b}$ decays.
The systematic errors of the b-tagging are mainly due to well understood
uncertainties in the decay length resolution 
and in the b-lifetime.
This selection allows limits to be set in the small region
of large $m_{\nt_2}$ and low $m_{\nt_1}$ that is not accessible within
the framework of the MSSM.

\item[(D5.4)]
The acoplanarity angle determined as described for cut (A3) 
is required to be greater than $20^{\circ}$,
if the number of reconstructed jets, obtained using the
Durham algorithm with $y_{\mathrm cut} = 0.005$, is
larger than or equal to three. This cut reduces potentially
mismeasured three-jet events.
\end{itemize}

The numbers of events remaining after each cut are listed
in Table~\ref{tab:nevD}.
The efficiencies for $\nt_2 \nt_1$ events falling into 
category (D) are listed in Table~\ref{tab:effDntWv}\@.

%
\subsubsection{Combined efficiencies and backgrounds for $\nt_1 \nt_2$}

The net efficiency for each mass pair combination is obtained
by taking the sum of the efficiencies for 
categories (C) and (D) weighted by the
fraction of signal events falling into each category.
Overall efficiencies for $\nt_2 \nt_1$
events are given in Table~\ref{tab:efftotal}.
As shown in this table, the efficiencies are 29--43\% if 
$\Delta M_{0} \geq 20$~GeV and $\leq 70$~GeV.
The total background expected for this search, 
which is the sum of the background
contributions from categories (C) and (D), is 0.96~events.

\vspace{1cm}

\begin{table}[hb]
\centering
\begin{tabular}{|l||r||r||r|r|r|r||r|r|r|}
\hline
   &  data  & total &
$\qq (\gamma)$ & $\ellell (\gamma)$ &
\multicolumn{1}{|c|}{`$\gamma \gamma$'} &
\multicolumn{1}{c||}{4-f} &
\multicolumn{3}{c|}{$\nt_1 \nt_2$}  \\
\hline
$m_{\nt_2}+m_{\nt_1}$ &   &     &         &              &            &
        & \multicolumn{3}{|c|}{160GeV} \\
\hline
$m_{\nt_2}-m_{\nt_1}$ &   &     &         &              &            &
        &  &  &  \\
\multicolumn{1}{|c||}{(GeV)} &  &     &         &              &            &
        & \multicolumn{1}{c|}{10} &
          \multicolumn{1}{c|}{30} &
          \multicolumn{1}{c|}{70}  \\
\hline
\hline
no cuts
&      &       &       &       &       &       &1000 &1000 &1000 \\
\hline
 Presel.+(D) 
& 2744 &  2309. & 796. &  38.6 & 1347. &  127 & 476 & 801 & 816 \\ 
\hline
 Cut (D1)
&  642 &  608. &  270. &  8.36 &  243. &  86.7 & 417 & 659 & 664 \\
\hline
 Cut (D2)
&  238 &  240. &  164. &  6.14 &  1.25 &  68.8 & 192 & 622 & 652 \\ 
\hline
 Cut (D3)
&   35 &  35.9 &  3.89 & 0.22 & 0.30 &  31.5 & 188 & 577 & 553 \\
\hline
\hline
$\Mvis \leq $20 GeV  
&    2 & 0.65 & 0.00 & 0.006 & 0.22 & 0.43 & 188 & 312 &  25 \\
\hline
 Cut (D4)
&    0 & 0.15 & 0.00 & 0.002 & 0.08 & 0.06 & 163 & 234 &   5 \\
\hline
\hline
$\Mvis > $20 GeV 
&   33 &  35.3 &  3.89 & 0.21 & 0.08 &  31.1 &   0 & 265 & 528 \\
\hline
 Cut (D5.1)
&    5 &  4.45 & 0.14 & 0.05 & 0.08 &  4.17 &   0 & 265 & 522 \\
\hline
 Cut (D5.2)
&    3 &  1.66 & 0.02 & 0.02 & 0.08 &  1.54 &   0 & 259 & 514 \\
\hline
 Cut (D5.3)
&    0 & 0.56 & 0.02 & 0.01 & 0.08 & 0.44 &   0 & 259 & 392 \\
\hline
 Cut (D5.4)
&    0 & 0.47 & 0.02 & 0.01 & 0.00 & 0.44 &   0 & 248 & 390 \\
\hline
\hline
 Net
&    0 & 0.62 & 0.02 & 0.02 & 0.08 & 0.50 & 163 & 482 & 395 \\
\hline
\end{tabular}
\caption[]{
  \protect{\parbox[t]{12cm}{
The remaining numbers of events for the
various background processes normalized to $\lumi$~pb$^{-1}$
after each cut in category (D). Numbers for
three simulated event samples of $\nt_1 \nt_2$ with
$\nt_2 \ra \nt_1 \ellell$ or $ \nt_1 \qq$ are
also given.
}} }
\label{tab:nevD}
\end{table}

\clearpage

\begin{table}[htb]
\centering
\begin{tabular}{|l||r|r|r|r|r|r|r|r|}
\hline
$m_{\nt_2}+m_{\nt_1}$(GeV)
& 100 & 110 & 120 & 130 & 140 & 150 & 160 &170  \\
\hline
\hline
$\Delta M_{0}$  &    &    &    &    &    &    &     &    \\
\hline
 3.0~GeV & 10 & 10 &  0 &  4 &  0 &  0 &  0 &  0 \\
\hline
 5.0~GeV & 21 & 16 & 14 &  9 &  5 &  4 &  3 &  0 \\
\hline
 10~GeV  & 37 & 38 & 40 & 37 & 40 & 40 & 31 & 33 \\
\hline
 20~GeV  & 47 & 52 & 53 & 50 & 58 & 57 & 59 & 60 \\
\hline
 30~GeV  & 51 & -- & 55 & -- & 57 & -- & 58 & -- \\
\hline
 50~GeV  & -- & 55 & -- & 61 & -- & 61 & -- & 57 \\
\hline
 70~GeV  & 45 & -- & 42 & -- & 43 & -- & 46 & -- \\
\hline
 80~GeV  & 34 & 30 & 31 & 32 & 30 & 28 & 30 & 32 \\
\hline
 90~GeV  & -- & 18 & -- & 15 & -- & 11 & -- & 11 \\
\hline
 110~GeV &    & -- &  8 & -- &  9 & -- &  8 & -- \\
\hline
 130~GeV &    &    &    & -- & -- &  6 & -- &  7 \\
\hline
 150~GeV &    &    &    &    &    & -- &  7 & -- \\
\hline
\end{tabular}
\caption[]{
  \protect{\parbox[t]{12cm}{
The detection efficiencies in percent for $\nt_1 \nt_2$
with $\nt_2 \ra \nt_1 \Zrv$ normalized to the number of
events in category (D).
}} }
\label{tab:effDntWv}
\end{table}

\begin{table}[htb]
\centering
\begin{tabular}{|l||r|r|r|r|r|r|r|r|}
\hline
$m_{\nt_2}+m_{\nt_1}$(GeV)
& 100 & 110 & 120 & 130 & 140 & 150 & 160 &170  \\
\hline
\hline
$\Delta M_{0}$  &    &    &    &    &    &    &     &    \\
\hline
 3.0~GeV &  1 &  1 &  0 &  0 &  0 &  0 &  0 &  0 \\
\hline
 5.0~GeV &  8 &  7 &  6 &  5 &  3 &  3 &  1 &  1 \\
\hline
 10~GeV  & 20 & 20 & 23 & 21 & 22 & 22 & 18 & 20 \\
\hline
 20~GeV  & 31 & 33 & 35 & 34 & 38 & 38 & 39 & 40 \\
\hline
 30~GeV  & 34 & -- & 38 & -- & 39 & -- & 42 & -- \\
\hline
 50~GeV  & -- & 39 & -- & 41 & -- & 43 & -- & 40 \\
\hline
 70~GeV  & 32 & -- & 29 & -- & 30 & -- & 32 & -- \\
\hline
 80~GeV  & 24 & 21 & 22 & 22 & 20 & 20 & 21 & 22 \\
\hline
 90~GeV  & -- & 13 & -- & 11 & -- &  8 & -- &  7 \\
\hline
 110~GeV &    & -- &  6 & -- &  6 & -- &  6 & -- \\
\hline
 130~GeV &    &    &    & -- & -- &  4 & -- &  5 \\
\hline
 150~GeV &    &    &    &    &    & -- &  4 & -- \\
\hline

\end{tabular}
\caption[]{
  \protect{\parbox[t]{12cm}{
The detection efficiencies in percent for $\nt_1 \nt_2$
with $\nt_2 \ra \nt_1 \Zv$ decay
for category (C) and (D) combined.
The invisible decay $\nt_2 \ra \nt_1\nunu$ which could occur
20\% of the time is assumed to be undetectable.
}} }
\label{tab:efftotal}
\end{table}

\subsection{Detection of chargino events with WW-like signature}
\label{sec:WWevents}

A special analysis for the chargino search was performed for the case
in which the chargino mass is
close to the W mass and the light neutralinos are almost massless. 
The event topology of such chargino pair events is  
similar to that of ordinary $\WW$ events but with 
somewhat larger missing energy as the neutralinos tend to have 
small but significant momentum.
If $\tan \beta \approx 1$ and $M_2 \approx \mu \approx 0$, the chargino 
and light neutralinos satisfy these conditions as described in 
Section 1.

This analysis may be sensitive to the details of the energy flow, especially in
the $\ee \ra \WW$ background simulation, as will become apparent in cuts
(E4) and (F5) in the selections that are presented in the following.
Therefore the observed energies in the simulated events
are scaled with the ratio of the centre-of-mass energies  
of the data and the simulated events,
when calculating efficiencies of background events 
at each energy (170.3~GeV and 172.3~GeV).

The search for such chargino events 
is performed by dividing the event sample
into two categories described as follows: 

 \subsubsection{Analysis E (${\boldmath N_{\mathrm ch} > 4}$  
and no isolated leptons)}
When both charginos decay into $\nt_i \qq'$ ($i=1,2$), 
the event shape is similar to 
that of the $\WW$ events in which both W's decay hadronically. 
For events with more than four charged tracks
($N_{\mathrm ch} > 4$) and with no isolated leptons (category (A))
the following cuts are applied: 
\begin{itemize}
\item[(E1)]
  The visible energy in the region defined by $|\cos \theta|>0.9$
  should be less than 20\% of the total visible energy.
  In addition, to reduce background from two-photon processes  
  $\cosmiss$ should be smaller than 0.9.
\item[(E2)]
  The magnitude of the momentum component longitudinal to the beam axis 
should be smaller than 25~GeV. This cut reduces the background
from $\ell\ell\qq$ final states.
\item[(E3)]
  The maximum EM cluster energy should be smaller than 35~GeV.  
\end{itemize}
\noindent
The above three cuts reject $\Zg$ events with a hadronic Z decay.
Since the signal events have missing energy and missing momentum due to
the invisible neutralinos,  
the following cut is applied to reduce the background from W-pair
events.
\begin{itemize}
\item[(E4)]
  The visible energy should be between 50 and 150 GeV.
  The distribution of the visible energy
  is shown in Fig.~\ref{fige1} after cut (E3).  
\end{itemize}
\noindent
The following two cuts are applied to reduce 
the contribution from $\ee \ra \qq$ events.
\begin{itemize}
\item[(E5)]
  The number of jets reconstructed  
  with the Durham algorithm using  
  a jet resolution parameter of $y_{\mathrm cut} = 0.005$
  should be at least four. 
\item[(E6)]
  The sum of the two highest jet energies ($E_1 + E_2$) should be 
  smaller than 100~GeV.
\end{itemize}

In Table~\ref{tab:nevE}, the remaining numbers of simulated events
for background processes and
for three samples of simulated $\chp_1 \chm_1$ events are given.

The main background comes from four-fermion processes 
as well as $\qq (\gamma)$ events.
Seven events are observed which is consistent with the 
total expected background of 7.0 events.
The net detection efficiency for chargino events is typically 18--20\% for 
an 80~GeV chargino decaying into a light stable 
neutralino ($m_{\nt_1} < 10$~GeV).
The efficiency does not drop by more than 1\%
when both charginos decay into $\Wrv \nt_2$ and $\nt_2$ decays 
into $\nt_1 \gamma$.

\begin{table}[htb]
\centering
\begin{tabular}{|l||r||r||r|r|r|r||r|r|r|}
\hline
   & data & total &
$\qq (\gamma)$ & $\ellell (\gamma)$ &
\multicolumn{1}{c}{`$\gamma \gamma$'} &
\multicolumn{1}{|c||}{4-f} &
\multicolumn{3}{c|}{$\chp_1 \chm_1$}  \\
\hline
$m_{\ch_1}$ (GeV)&   &     &         &                    &            &
        & \multicolumn{1}{c|}{80} & 
          \multicolumn{1}{c|}{80} &
          \multicolumn{1}{c|}{80}  \\
$m_{\nt_1}$ (GeV)&   &    &         &                    &            &
        & \multicolumn{1}{c|}{0} & 
          \multicolumn{1}{c|}{0} & 
          \multicolumn{1}{c|}{10}  \\
$m_{\nt_2}$ (GeV)&   &    &         &                    &            &
        & \multicolumn{1}{c|}{--} & 
          \multicolumn{1}{c|}{2} & 
          \multicolumn{1}{c|}{--}  \\
\hline
\hline
no cuts
&  &  &     &      &        &       & 1000  & 1000  & 1000 \\
\hline
 Presel.+(E)
& 1448 & 1282. & 777. & 13.7  &   408. & 83.0 & 522 & 510 & 499 \\
\hline
 Cut (E1)
&  450 & 429.  & 271. & 2.54  &   96.1 & 59.0 & 412 & 410 & 400 \\
\hline
 Cut (E2)
&  418 & 408.  & 255. & 2.11  &   96.0 & 54.7 & 356 & 365 & 326 \\
\hline
 Cut (E3)
&  281 & 270.  & 123. & 0.63 &  96.0 & 50.3 & 348 & 355 & 320 \\
\hline
 Cut (E4)
&  39  & 41.8  & 28.4 & 0.47 & 0.59 & 12.3 & 292 & 258 & 310 \\
\hline
 Cut (E5)
&   7 &  7.96  & 3.22 & 0.000 & 0.08 & 4.66 & 200 & 194 & 184 \\
\hline
 Cut (E6)
&   7 &  7.00  & 2.44 & 0.000 & 0.08 & 4.47 & 199 & 186 & 183 \\
\hline
\end{tabular}
\caption[]{
  \protect{\parbox[t]{12cm}{
The remaining numbers of events for the 
various background processes normalized to 1.0 and 9.3~pb$^{-1}$ 
for $\sqrt{s}$ = 170.3 and 172.3~GeV, respectively,
are compared after each cut in category (E).
Numbers for
three simulated event samples of $\chp_1 \chm_1$ with
$\chp_1 \ra (\nt_1$ or $\nt_2) {\mathrm{W}}^*$ are
also given (starting from 1000 events).
} }}
\label{tab:nevE}
\end{table}

\subsubsection{Analysis F (${\boldmath N_{\mathrm ch} > 4}$  
with isolated leptons)}
When one of the charginos decays into $\nt_i \qq'$ ($i=1,2$)
and the other into $\nt_j \ell \nu$ ($j=1,2$), the event shape is similar to 
that of $\WW \ra \nulqq$ events. 
Events with more than four charged tracks ($N_{\mathrm ch} > 4$) and 
at least one isolated lepton (category (B)) 
are selected and the following cuts are applied: 
\begin{itemize}
\item[(F1)]
  To reduce the contribution of 
  events from two-photon processes and $\Zg$ events
  where the $\gamma$ escaped undetected down the beam pipe,
  $\cosmiss$ should be smaller than 0.9.
\item[(F2)]
  To reduce the background from two-photon processes,
  $P_t$ and $P_t^{\mathrm{HCAL}}$ were required to be
  greater than 10~GeV.
\item[(F3)]
  The visible energy should be smaller than 125~GeV.
  This cut is effective in rejecting the well contained 
  $\ee \ra \qq$ and W pair events. 
\item[(F4)]
  The isolated lepton energy should be smaller than 50~GeV. 
\item[(F5)]
  The invariant mass of the event excluding the highest momentum isolated 
  lepton should be between 15 and 80~GeV. 
  The $M_{\mathrm rest}$ distribution is shown in Fig.~\ref{figf1}
  for the data, the expected background processes and the signal after
  cut (F4).
\end{itemize}

In Table~\ref{tab:nevF}, the remaining numbers of simulated events
for background processes and
for three samples of simulated $\chp_1 \chm_1$ events are given.

\begin{table}[htb]
\centering
\begin{tabular}{|l||r||r||r|r|r|r||r|r|r|}
\hline
   & data & total &
$\qq (\gamma)$ & $\ellell (\gamma)$ &
\multicolumn{1}{c}{`$\gamma \gamma$'} &
\multicolumn{1}{|c||}{4-f} &
\multicolumn{3}{c|}{$\chp_1 \chm_1$}  \\
\hline
$m_{\ch_1}$ (GeV)&   &     &         &                    &            &
        & \multicolumn{1}{c|}{80} & 
          \multicolumn{1}{c|}{80} &
          \multicolumn{1}{c|}{80}  \\
$m_{\nt_1}$ (GeV)&   &    &         &                    &            &
        & \multicolumn{1}{c|}{0} & 
          \multicolumn{1}{c|}{0} & 
          \multicolumn{1}{c|}{10}  \\
$m_{\nt_2}$ (GeV)&   &    &         &                    &            &
        & \multicolumn{1}{c|}{--} & 
          \multicolumn{1}{c|}{2} & 
          \multicolumn{1}{c|}{--}  \\
\hline
\hline
no cuts
&  &  &     &      &        &       & 1000  & 1000  & 1000 \\
\hline
 Presel.+(F)
& 1296 & 1027. &  18.9 &  24.9 &  939. &  44.0 & 324 & 312 & 327 \\
\hline
 Cut (F1)     
&  642 &  601. &  7.79 &  10.5 &  546. &  36.3 & 301 & 293 & 294 \\
\hline
 Cut (F2)
&   37 &  39.1 &  2.28 &  3.22 & 0.10 &  33.5 & 284 & 268 & 281 \\
\hline
 Cut (F3)
&   16 &  16.3 & 0.30 &  1.39 & 0.10 &  14.5 & 266 & 218 & 274 \\
\hline
 Cut (F4)
&   14 &  15.5 & 0.30 &  1.24 & 0.05 &  13.9 & 252 & 207 & 268 \\
\hline
 Cut (F5)
&    8 &  10.4 & 0.22 & 0.31 & 0.00 &  9.83 & 235 & 167 & 238 \\
\hline
\end{tabular}
\caption[]{
  \protect{\parbox[t]{12cm}{
The remaining numbers of events for the 
various background processes normalized to 1.0 and 9.3~pb$^{-1}$ 
for $\sqrt{s}$ = 170.3 and 172.3~GeV, respectively,
are compared after each cut in category (F).
Numbers for
three simulated event samples of $\chp_1 \chm_1$ with
$\chp_1 \ra (\nt_1$ or $\nt_2) {\mathrm{W}}^*$ are
also given (starting from 1000 events).
} }}
\label{tab:nevF}
\end{table}
The main background comes from four-fermion processes.
Eight events are observed which is consistent with the 
total expected background of 10.4 events.
The net detection efficiency for chargino events is typically 24\% for 
a 80~GeV chargino decaying into a light 
stable neutralino ($m_{\nt_1} < 10$~GeV).
The efficiency drops by about 7\%
when both charginos decay into $\Wrv \nt_2$ and $\nt_2$ decays 
into $\nt_1 \gamma$.

\subsubsection{Combined efficiencies and backgrounds for 
analyses (E) and (F)}
The overall efficiency for each mass pair combination is obtained
by taking the weighted sum of the efficiencies for categories (E) and (F). 
Overall efficiencies for $\chp_1 \chm_1$
events are 35--45\%. 
The total number of expected background events is 17.4, while 15 events are 
observed.  

\subsection{Systematic errors and corrections }
In analyses (A)-(D),
systematic errors on the number of expected signal events arise from 
the following sources:
the measurement of the integrated luminosity (0.6\%),
Monte Carlo statistics in the various signal samples and the
interpolation errors of the efficiencies at arbitrary values
of $m_{\chp_1}$ ($m_{\nt_2}$) and $m_{\nt_1}$ (2--10\%),
modelling of the cut variables in the Monte Carlo
simulations\footnote{
This is estiimated by comparing the efficiencies obtained by shifting
each cut variable by the maximal possible shift in the corresponding
distribution which still gives agreement between data and Monte
Carlo.} 
(2--4\%), errors
due to fragmentation uncertainties in hadronic decays ($< 2$\%),
the matrix elements leading to 
different decay parameters ($< 5$\%)
and effects of detector calibration ($< 1$\%).
The effect of possible trigger inefficiencies
has been checked and found to be negligible.
These systematic errors are considered to be independent and are added
in quadrature (7--12\%).

In analyses (A)-(D),
the systematic errors on the expected number of background
events used to obtain the limits (when combining present with 
previous results) are due to:
Monte Carlo statistics in the simulated background events,
uncertainties in the amount of two-photon background, estimated by
fitting the $P_t$ distributions of simulated two-photon events
and the data (30\%); and uncertainties in the simulation of
the four-fermion processes, which are estimated by taking the difference
between the predictions of
the grc4f~\cite{grc4f} and the EXCALIBUR~\cite{excalibur} generators 
(17\% for chargino, 20\% for neutralino). The systematic errors due to
the modelling of the cut variables in the detector simulation are 12\%
for the two-photon processes and 4\% for the four-fermion processes in
the case of 
the chargino selection. In the neutralino search these errors become
14\% and 5\%, respectively. 
For the neutralino selection
an additional uncertainty of 11\% due to the b-tagging is included.
A 100\% systematic error is assumed for the number of background events
expected for the $\qq (\gamma)$ and $\ellell (\gamma)$ states.
Therefore, the total expected number of background events is
estimated for the chargino search to be
$0.87 \pm 0.19$ ($0.02 \pm 0.02$ from $\qq(\gamma)$,
$0.03\pm0.03$ from $\ellell(\gamma)$, $0.34\pm0.17$ from two-photon 
processes and $0.48\pm 0.09$ from four-fermion final states),
and for the neutralino search  
$0.96 \pm 0.22$ ($0.02 \pm 0.02$ from $\qq(\gamma)$,
$0.02\pm0.02$ from $\ellell(\gamma)$, $0.34\pm0.18$ from two-photon 
processes and $0.58\pm 0.13$ from four-fermion final states).

In analyses (E) and (F),
systematic errors on the number of expected signal events arise
from the following sources:
the measurement of the integrated luminosity (0.6\%);
four jet selection in analysis E (0.9\%); 
Monte Carlo statistics of the signal samples and interpolation of the
efficiencies at arbitrary values of $m_{\chp_i}$ and $m_{\nt_j}$ (4-5\%);
preselection (mainly the cut on the energy deposited in the 
forward calorimeter) ($< 2$\%);
fragmentation uncertainties in hadronic decays (2\%);
and the energy scale (0.6\%).

In analyses (E) and (F)
the systematic errors on the number of expected four-fermion 
background events (dominated by W-pair production)
come from modelling of the processes (2.4\%), estimated 
by comparing the numbers obtained with 
the grc4f, EXCALIBUR
and PYTHIA~\cite{PYTHIA} generators,
calibration of the energy scale (6\%), and the beam energy uncertainty (0.3\%).
The error due to the uncertainty on the W mass~\cite{PDG} is 0.4\%.
The systematic errors on the number of expected $\ee \ra \qq (\gamma)$
background events is due to 
the energy scale (7\%), and to the modelling of the process (11\%),
estimated by comparing the numbers obtained with
the PYTHIA and HERWIG \cite{HERWIG} generators. 
The error on the modelling of the processes is mainly due to the
limited statistics of the selected Monte Carlo events. 
The combined error for four-fermion events
and for $\ee \ra \qq (\gamma)$ events 
due to preselection (mainly the cut on the 
energy deposited in the forward calorimeter) is $< 1.6$\% and 
the one due to the four jet selection in analysis E is 1.4\%. 
A 100\% systematic error was assigned to the number of lepton-pair 
and two-photon background events.
The errors due to the energy scale for
four-fermion and for $\ee \ra \qq (\gamma)$ background events 
were added linearly.
All the other errors were added quadratically. 
The total number of background events was estimated to be $17.45\pm1.29$.
The total number of events surviving all cuts was 15. 

The rate of events in which the measured energy  in the
SW, FD or GC calorimeters, 
due to noise and beam related background, 
exceeded the thresholds in the preselection
is about 2.3\%  as estimated
from random beam crossing events. Since this effect is not modelled in
the simulation, this effect is taken into account by scaling
the detection efficiencies by this amount.
This correction is applied for all the analyses (A)-(F). 

The systematic errors on the numbers of expected signal and background
events are summarized in Tables~\ref{tab:systsig} and
\ref{tab:systbkgd} respectively.

\begin{table}[htb]
\centering
\begin{tabular}{|l||c|c|c|c|c|c|}
\hline
        & Analyses (A) to (D) & Analyses (E) and (F) \\
\hline
\hline
Integrated luminosity & 0.6   \% & 0.6  \% \\
MC statistics         & 2$-$10  \% & 4$-$5  \% \\
Cut variables         & 2$-$4   \% &   $-$     \\
Fragmentation         & 2     \% & 2    \% \\
Matrix element        & 5     \% &   $-$     \\
Detector calibration  & 1     \% &   $-$     \\
4-jet selection (E)   &   $-$      & 0.9  \% \\
Preselection          &   $-$      & 2    \% \\
Energy scale          &   $-$      & 0.6  \% \\
\hline
\end{tabular}
\caption[]{
  \protect{\parbox[t]{12cm}{
Systematic uncertainties on the numbers of expected signal events.
} }}
\label{tab:systsig}
\end{table}
\begin{table}[htb]
\centering
\begin{tabular}{|l|l|c|c|c|c|c|c|c|c|}
\hline
 & & 4-f & $\qq$ & $\gamma\gamma$ & $\ellell$ \\ 
\hline
\hline
Chargino search: 
          & Modelling     & 17 \%   &   $-$    & 30 \% &   $-$    \\
Analyses (A) to (C)
          & Cut variables &  4 \%   &    $-$   & 12 \% &   $-$    \\
          & others        &         & 100 \%  &       & 100 \% \\
\hline
Neutralino search: 
          & Modelling     & 20 \%   &   $-$     & 30 \% &   $-$    \\
Analyses (B) and (D)
          & Cut variables & 5 \%    &    $-$    & 14 \% &   $-$    \\
          & b tagging     & 11 \%   &    $-$    & 11 \% &   $-$    \\
          & others        &         & 100 \%  &       & 100 \% \\
\hline
          & Modelling     & 2.4 \%  & 11 \%   &  $-$    &   $-$    \\
          & Energy scale  & 6 \%    & 7 \%    &  $-$    &   $-$    \\
          & Beam energy   & 0.3 \%  &    $-$    &  $-$    &   $-$    \\
Analyses (E) and (F)
          & W mass        & 0.4 \%  &    $-$    &  $-$    &   $-$    \\
          & Preselection  & 1.6 \%  & 1.6 \%  &  $-$    &   $-$    \\
    & 4-jet selection (E) & 1.4 \%  & 1.4 \%  &  $-$    &   $-$    \\
          & others        &         &         & 100\% & 100 \% \\
\hline
\end{tabular}
\caption[]{
  \protect{\parbox[t]{12cm}{
Systematic uncertainties on the numbers of expected background events.
} }}
\label{tab:systbkgd}
\end{table}

\vspace{2cm}
\section{Results}
\subsection{Limits on the $\chp_1 \chm_1$ and $\nt_1 \nt_2$ cross-sections}

A model-independent interpretation is formed by calculating the
95\% confidence level (C.L.) upper limits on the production
cross-sections for $\chp_1 \chm_1$ and $\nt_1 \nt_2$ assuming
the specific decay modes 
$\chpm_1 \ra \nt_1 \Wvpm$ and $\nt_2 \ra \nt_1 \Zv$.
From the observation of no events at $\roots = 170 - 172$~GeV
in analyses (A)-(D), 
and using the signal detection efficiencies and their uncertainties,
exclusion regions
are determined using the procedure outlined in \cite{PDG}, and
incorporating systematic errors following the method given in
\cite{Cousins} by numerical integration, assuming Gaussian errors.
To compute the 95\% C.L. upper limits, the previous results obtained
at $\roots = 161$~GeV~\cite{LEP16-opal} (including the observed
candidates as well as background expectations where kinematically
allowed) have been combined with the present results.
For the combination of results from different energies it is assumed that
the cross-sections are proportional to $\tilde \beta /s$,
where $\tilde \beta$ is the momentum of the final state $\chp_1$
or $\nt_2$ in the centre-of-mass system normalized to the beam energy.

Contours of the upper limits for the $\chp_1 \chm_1$
cross-sections are shown in Fig.~\ref{figsum1}
assuming $\chpm_1 \ra \nt_1 {\mathrm W}^{*\pm}$
with 100\% branching fraction.
Similarly, contours of the upper limit for the $\nt_1 \nt_2$
cross-sections are shown assuming 100\% branching fraction
for $\nt_2 \ra  \nt_1 {\mathrm Z}^*$ (Fig.~\ref{figsum2}).
The Standard Model branching fractions are used for
the ${\mathrm W}^{*}$ and ${\mathrm Z}^*$ decays,
including the invisible decay mode ${\mathrm Z}^* \ra \nunu$
and taking into account phase-space effects for decays
into heavy particles (especially ${\mathrm b} \bar{\mathrm b}$).
Although these limits do not depend on the details of the SUSY models
considered, a ``typical" field
content\footnote{
These field contents arise from the MSSM parameters necessary to
give as close as possible gaugino masses as those being considered
with some mixture of $\tilde{\mathrm W}^{\pm}$ and
$\tilde{\mathrm H}^{\pm}$ for charginos and
$\tilde{\gamma}$, $\tilde{\mathrm Z}$, and $\tilde{\mathrm H}^0_{j}$
for neutralinos.  When more than one set of  parameters lead to the
same set of gaugino masses, the set not yet excluded or with
more moderate values of $|\mu|$ and $M_2$ was chosen.
}
of the gauginos is assumed, leading to particular 
production angular distributions that are subsequently used
in the estimation of detection efficiencies.
There are differences in detection efficiencies
due to variations in the angular distributions resulting
from using different MSSM
parameters corresponding to the same mass combination.
The variation of the efficiency is observed to be $<2$\%.
Of the parameters examined, those which result in
the lowest efficiency are used.

If the cross-section for $\chp_1 \chm_1$ is larger than
3.0~pb and $\Delta M_+$ is larger than 5~GeV,
it is possible to exclude at 95\% C.L. the $\chp_1$ up to
the kinematic limit for $\chp_1$ decay via W$^{*}$.
This is achieved by using analyses (A) to (C) in the non-hatched
regions of Fig.~\ref{figsum1} and using analyses (E) and (F) in the
hatched region.
Furthermore $\nt_2$ masses up to the kinematical
boundary  of ($m_{\nt_1} + m_{\nt_2}) < \roots$ are excluded
at 95\% C.L. for $10 \le \Delta M_0 \le 80$~GeV,
if the cross-section for $\nt_1 \nt_2$ is larger than
2.0~pb.

\subsection{Limits in the MSSM parameter space}
\label{interp}
The results of the above searches can be interpreted within the
framework of the constrained MSSM.
The phenomenology of the gaugino-higgsino
sector of the MSSM is mostly determined by the 
parameters $M_2$, $\mu$ and $\tan\beta$ defined earlier.
In the absence of light sfermions and light SUSY Higgs
particles, these three parameters are sufficient to describe
the chargino and neutralino sector completely.  
Within the constrained MSSM (CMSSM), a large value of the common 
scalar mass, $m_0$ (e.g., $m_0 = 1$~TeV) leads to heavy sfermions
and therefore to a negligible 
suppression of the cross-section due to interference 
from $t$-channel sneutrino exchange, and chargino and
neutralino decays would proceed
via a virtual $\mathrm{W}^*$ or $\mathrm{Z}^*$, respectively.
On the other hand, a light $m_0$ results in low values of
the masses of the $\snu$ and $\sell$,
thereby enhancing the contribution of the $t$-channel exchange diagrams 
that may
have destructive interference with $s$-channel diagrams,
thus reducing the cross-section for chargino pair production.
Small values of $m_0$ also tend to enhance the leptonic branching
ratio of charginos, often leading to smaller detection efficiencies.
For neutralino pair production the $t$-channel selectron exchange diagram
may interfere positively with the $s$-channel Z boson diagram to
enhance this cross-section, but the size of the chargino cross-section
almost always remains larger.
The results are therefore presented for two cases: $m_0 = 1$~TeV and the 
smallest value of
$m_0$ that is compatible with current limits on the
$\tilde{\nu}$ mass 
($m_{\snu_L} > 43$~GeV~\cite{PDG}), and
OPAL limits on the $\tilde{\ell}$ mass, particularly
right-handed smuon and selectron pair 
production~\cite{LEP16-slepton}.  
This latter ``minimum $m_0$" case gives the lowest $\chp_1 \chm_1$
production cross-section for $\tan \beta = 1.0$ but not necessarily
for larger $\tan \beta$ values.

In the region of 
small $M_2$ and large $\mu$, the $\chpm_1$ and $\nt_1$ would be 
almost pure gauginos, resulting in large
$\chp_1 \chm_1$ cross-sections, but small
$\nt_1 \nt_2$ cross-sections.
They would be mostly higgsino 
for large $M_2$ and small $\mu$, and in this case, the cross-section
for $\nt_1 \nt_2$ is such that the neutralino-specific 
searches can contribute significantly.

From the input parameters $M_2$, $\mu$, $\tan \beta$,
$m_0$ and $A$ (the trilinear coupling),
masses, production cross-sections and
branching fractions are calculated
according to the CMSSM~\cite{Bartl,chargino-theory,Ambrosanio,Carena}. 
For each set of input parameters, the total number of
$\chp_1 \chm_1$ (for analyses (A)-(C)),
$\nt_1 \nt_2$ and $\nt_1 \nt_3$ (for analyses (C)-(D))
$\chp_1 \chm_1$, $\chpm_1 \chmp_2$ and $\chp_2 \chm_2$ (for analyses (E)-(F)) 
events
expected to be observed is found using
the known integrated luminosity, calculated cross-sections,
branching ratios, and 
the detector efficiencies which depend upon the masses of these
particles.
The efficiency for detecting
$\nt_1 \nt_3$ events, even for decays through SUSY Higgs bosons,
is found to be greater than for
$\nt_1 \nt_2$ events, so
the efficiency functions for $\nt_1 \nt_2$ were used to obtain
conservative limits.
The decay involving $\nt_{2,3} \ra \tilde{\ell} \ell$ when
$m_{\nt_{2,3}} > m_{\tilde{\ell}}$
and the decay $\nt_{2,3} \ra \nt_1 \gamma$ are
assumed to be undetectable in this analysis.

Slepton and sneutrino masses,
cross-sections, and branching ratios are also 
determined at each set of CMSSM parameters.  
When the minimum $m_0$ case is considered, the value of
$m_0$ is decreased until it is just consistent with both
$\tilde{\nu}$ mass limits~\cite{PDG} and
the 95\% C.L. upper limits on the product of cross-section
and BR$^{2}(\tilde{\ell} \rightarrow \ell \nt_1)$ for
$ \tilde{\mu}_R^+ \tilde{\mu}_R^-$
and  
$ \tilde{\mathrm{e}}_R^+ \tilde{\mathrm{e}}_R^-$
pair production as given in Ref.~\cite{LEP16-slepton}.

The following regions of the CMSSM parameters are scanned:
$0\le M_2 \le 1500$~GeV,
$|\mu| \le 500$~GeV, and
$A = \pm M_2, \,\, \pm m_0$ and 0.
The typical scan step is 0.2~GeV.
It has been checked that the scanned ranges of parameters are
large enough that the exclusion regions change
negligibly for larger ranges.
No significant dependence on $A$ is observed.

Using the results of analyses (A)-(D),
the 95\% C.L. upper limit on the expected number of 
events is determined as described previously.
Figure~\ref{fig_mssm} 
shows the resulting exclusion regions 
(shaded regions) in the ($M_2$,$\mu$) plane
for $\tan\beta = 1.5$ and 35.
The region of $M_2$-$\mu$ excluded is enlarged significantly with
respect to the results at $\sqrt{s}=161$~GeV alone~\cite{LEP16-opal}.
In the CMSSM the gauginos
have a common mass, $m_{1/2}$, at the GUT scale, therefore
the gluino mass ($M_3 \equiv m_{\tilde{g}}$)
is directly related to $M_2$ by
$M_3/M_2=\alpha_s/\frac{\alpha}{{\mathrm sin}^2 \theta_W}$\ .
Figure~\ref{fig_mssm} therefore includes a scale indicating the
corresponding mass limits for gluinos.

The problem associated with $\tan\beta$ approaching 1.0
is clear from
Fig.~\ref{label_12} where the region near $M_2 = \mu = 0$
is not excluded 
by results from analyses (A)-(D), or
from the $\mathrm{Z}^0$ width and
direct neutralino searches at LEP1~\cite{LEP1-neut}.
For the case of $\tan\beta = 1.0$, the
results from analyses (E) and (F) are therefore invoked, and the
consequent excluded regions are shown as the hatched regions
overlaid onto the regions excluded by analyses (A)-(D)
in Fig.~\ref{label_12}(a) and (b).
The areas near $M_2 = \mu = 0$ not excluded by previous analyses are now
excluded.

The restrictions on the CMSSM parameter space presented can be
transformed into exclusion regions in ($m_{\nt_1}$,$m_{\chpm_1}$) or
($m_{\nt_2}$,$m_{\nt_1}$) mass space.
A given mass pair is considered excluded only if
{\it all} CMSSM parameters in the scan which lead
to the same values of mass pairs being considered are
excluded at the 95\%~C.L.
In the ($m_{\nt_1}$,$m_{\chpm_1}$) plane, Fig.~\ref{masslimc}
shows the corresponding
95\% C.L. exclusion regions for $\tan\beta = 1.0$, 1.5 and 35.
The region extending beyond the kinematic limit for 
chargino pair production in Fig.~\ref{masslimc}(a) is due
to the direct topological search for neutralinos at 
LEP1~\cite{LEP1-neut} illustrated in Fig.~\ref{label_12}.
The analogous exclusion regions in the 
($m_{\nt_1}$,$m_{\nt_2}$) plane are
shown in Fig.~\ref{masslimn}.
A smaller fraction of the accessible region of mass space for 
neutralino production
is excluded because of the smaller predicted cross-sections
for neutralinos.
The portion of the excluded region extending beyond the kinematic limit
$(m_{\nt_1}+m_{\nt_2}) = 172\mathrm~GeV$ is due to the exclusion of chargino
production for the relevant CMSSM parameters.
The lower limits of the chargino and neutralino masses are listed
in Table~\ref{tab:results}.


A similar procedure is followed for other values of $\tan\beta$ to
find the lower limit on the mass of the $\nt_1$ as
a function of $\tan\beta$ for $m_0 = 1$~TeV and minimum $m_0$
with the result shown in Fig.~\ref{mass_tanb}(a).
Each lower limit on $m_{\nt_1}$ in Fig.~\ref{mass_tanb}(a) corresponds to
a particular value of $M_2$, $\mu$, $\tan\beta$, and
minimum $m_0$ consistent with slepton mass limits.
Figure~\ref{mass_tanb}(b) shows the corresponding sneutrino mass for those
SUSY parameters where the lowest $\nt_1$ mass was found, as a function of
$\tan\beta$.

Returning to the assumption of gauge unification at the
GUT scale and the CMSSM, a limit on $M_2$ 
can be obtained as a function of $m_0$ for a given
value of $\mu$ and $\tan \beta$.  
Under this assumption, limits on gluino and squark masses are
then implied.
The average of the $\su_R$, $\su_L$, $\sd_R$ and $\sd_L$ masses  
($m_{\sq}$) can be calculated from $m_0$, $M_2$ and $\tan \beta$ 
(see equations (1)-(4)) in the CMSSM framework.
Limits in the ($m_{\sq}$,$m_{\sg}$) plane can therefore be
calculated and are shown in Fig.~\ref{squark1}. 
The limit of Fig.~\ref{squark1}(a) can be compared with the mass limit from 
current direct 
$\sq$ and $\sg$ searches at the Tevatron.
For $\tan \beta = 4$ and $\mu = -400$~GeV, a gluino mass limit 
of 270~GeV was obtained for the case of $m_{\sq} > 500$~GeV.
Under these assumptions
the limit is significantly better than those obtained 
from direct searches by  
the CDF and D0 collaborations at the Tevatron~\cite{CDFsgsq, DOsgsq}.
The CDF result is also shown in Fig.~\ref{squark1}(a).
In the region below the diagonal curve the lightest slepton 
or sneutrino mass becomes negative in the CMSSM framework.

\begin{table}[b]
\centering
\begin{tabular}{|c||c|c|c|c|c|c|}
\hline
  Mass  & \multicolumn{2}{c|}{$\tan \beta = 1.0$}
        & \multicolumn{2}{c|}{$\tan \beta = 1.5$}
        & \multicolumn{2}{c|}{$\tan \beta = 35$} \\
    \cline{2-7}
  GeV   & Min. $m_0$ & $m_0 = 1$~TeV
        & Min. $m_0$ & $m_0 = 1$~TeV
        & Min. $m_0$ & $m_0 = 1$~TeV  \\
\hline
\hline
$m_{\ch_1}$ & $>65.7$ & $>84.5$ & $>72.1$ & $>85.0$ & $>74.4$ & $>85.1$\\
$m_{\nt_1}$ & $>13.3$ & $>24.7$ & $>23.9$ & $>34.6$ & $>40.9$ & $>43.8$\\
$m_{\nt_2}$ & $>46.9$ & $>46.9$ & $>45.3$ & $>56.5$ & $>74.6$ & $>85.5$\\
$m_{\nt_3}$ & $>75.8$ & $>90.1$ & $>94.1$ & $>101.7$ & $>116.5$ &$>116.5$
\\
\hline
\end{tabular}
\caption[]{
   \protect{\parbox[t]{12cm}{
Lower limits at 95\% C.L. obtained
on the lightest chargino mass,
and the masses of the three lightest neutralinos, in GeV.
These limits are given for
$\Delta M_+ \geq 10\mathrm~GeV$ and $\Delta M_0 \geq 10\mathrm~GeV$.
Two cases are considered:
$m_0 = 1\mathrm~TeV$ and the smallest $m_0$ possible that complies
with the LEP1 $\snu$ and OPAL $\tilde{\ell}$ limits.
}} }
\label{tab:results}
\end{table}

\section{Summary and Conclusion}

A data sample corresponding to
an integrated luminosity of $\lumi$~pb$^{-1}$
at $\roots = $170 and 172~GeV
collected with the OPAL detector has been analysed
to search for pair production of charginos and neutralinos
predicted by supersymmetric theories.
The expected
background for each search is 0.9 events
and no candidate events are observed in either search.
For the case of $\tan \beta$ close to 1.0, and
$M_2 \approx \mu \approx 0$, a new search is performed and no evidence for
an excess of W-pair-like events is observed.
The exclusion limits on $\ch_1$ and $\nt_j$ production are significantly
higher with respect to the results obtained by OPAL at
$\roots = $130~GeV, 136~GeV~\cite{LEP15-opal} and 161~GeV~\cite{LEP16-opal}.
The 95\% C.L. lower mass limit of the chargino is close to
the kinematic limit within the framework of the
MSSM.
Assuming that the lightest  
chargino is heavier than the lightest neutralino by
more than 10~GeV, a lightest chargino mass limit at 
95\% C.L. of
84.5~GeV is obtained (for $\tan \beta \geq 1.0$) 
if the universal scalar mass, $m_0$,
is larger than 1~TeV. The mass limit is 65.7~GeV for the smallest
$m_0$ compatible with 
limits on sneutrino masses and OPAL limits on slepton masses.
The lower limit on the lightest neutralino mass ($m_{\nt_1}$) at 
95\% C.L. for $\tan \beta \geq 1.0$ is 24.7~GeV for $m_0 = 1$~TeV
and 
13.3~GeV for the minimum $m_0$ scenario.
These limits are also given as functions of $\tan\beta$ and 
are generally higher for larger values of $\tan\beta$.
If limits on $M_2$ are interpreted in the framework of the CMSSM,
limits on gluino and squark masses are complementary to 
those obtained from direct searches.

\par
\vspace*{1.cm}
\section*{Acknowledgements}
\noindent
We particularly wish to thank the SL Division for the efficient operation
of the LEP accelerator at all energies
 and for
their continuing close cooperation with
our experimental group.  We thank our colleagues from CEA, DAPNIA/SPP,
CE-Saclay for their efforts over the years on the time-of-flight and trigger
systems which we continue to use.  In addition to the support staff at our own
institutions we are pleased to acknowledge the  \\
Research Corporation, USA, \\
Department of Energy, USA, \\
National Science Foundation, USA, \\
Particle Physics and Astronomy Research Council, UK, \\
Natural Sciences and Engineering Research Council, Canada, \\
Israel Science Foundation, administered by the Israel
Academy of Science and Humanities, \\
Minerva Gesellschaft, \\
Benoziyo Center for High Energy Physics,\\
Japanese Ministry of Education, Science and Culture (the
Monbusho) and a grant under the Monbusho International
Science Research Program,\\
German Israeli Bi-national Science Foundation (GIF), \\
Bundesministerium f\"ur Bildung, Wissenschaft,
Forschung und Technologie, Germany, \\
National Research Council of Canada, \\
Hungarian Foundation for Scientific Research, OTKA T-016660, 
T023793 and OTKA F-023259.\\




\bibliographystyle{unsrt}


\newpage

\begin{figure}
\begin{minipage}{16.cm}
\begin{center}\mbox{
\epsfig{file=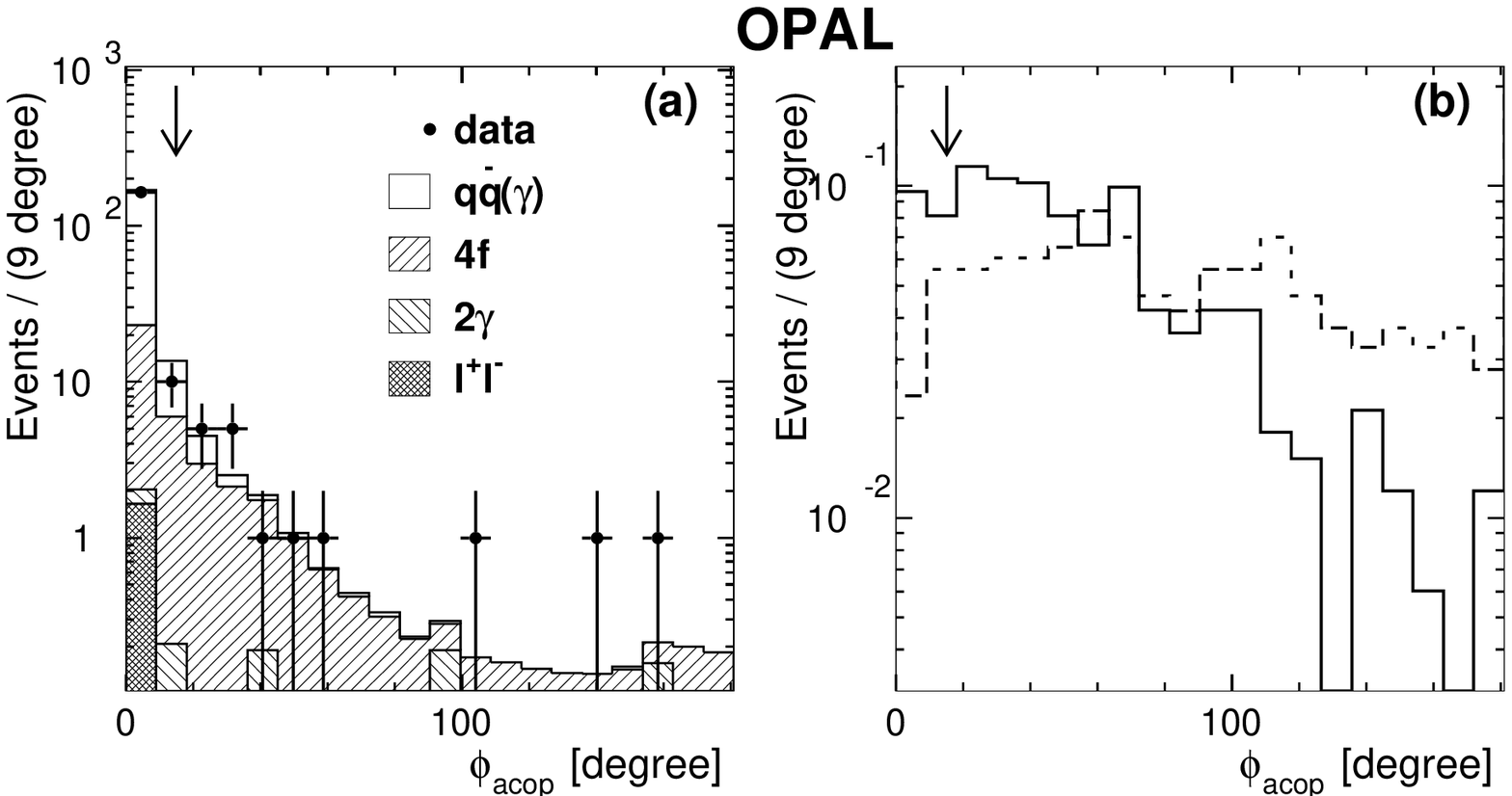,width=16.0cm}
}\end{center}
\caption[1]
{
The distributions of the acoplanarity angle, $\phiacop$, after cut (A2). 
In (a) are shown the predicted contributions
from background processes: 
dilepton events (double hatched area), 
two-photon processes (negative slope hatching area), 
four-fermion processes (including W-pair events) 
(positive slope hatched area),
and multihadronic events (open area). In each case the distribution
has been normalized to $\lumi$~pb$^{-1}$. Also shown in (a) is
the distribution of the data (dark circles). In (b) predictions from
simulated chargino events are shown for $m_{\ch_1}=80$~GeV and 
$m_{\nt_1}=60$~GeV (solid line histogram) and for
$m_{\ch_1}=80$~GeV and $m_{\nt_1}=20$~GeV (dotted line histogram).
The arrows shown indicate the position of the cut used in the
event selection.
}
\label{fig:cataacop}
\end{minipage}
\end{figure}

\begin{figure}
\begin{minipage}{16.cm}
\begin{center}\mbox{
\epsfig{file=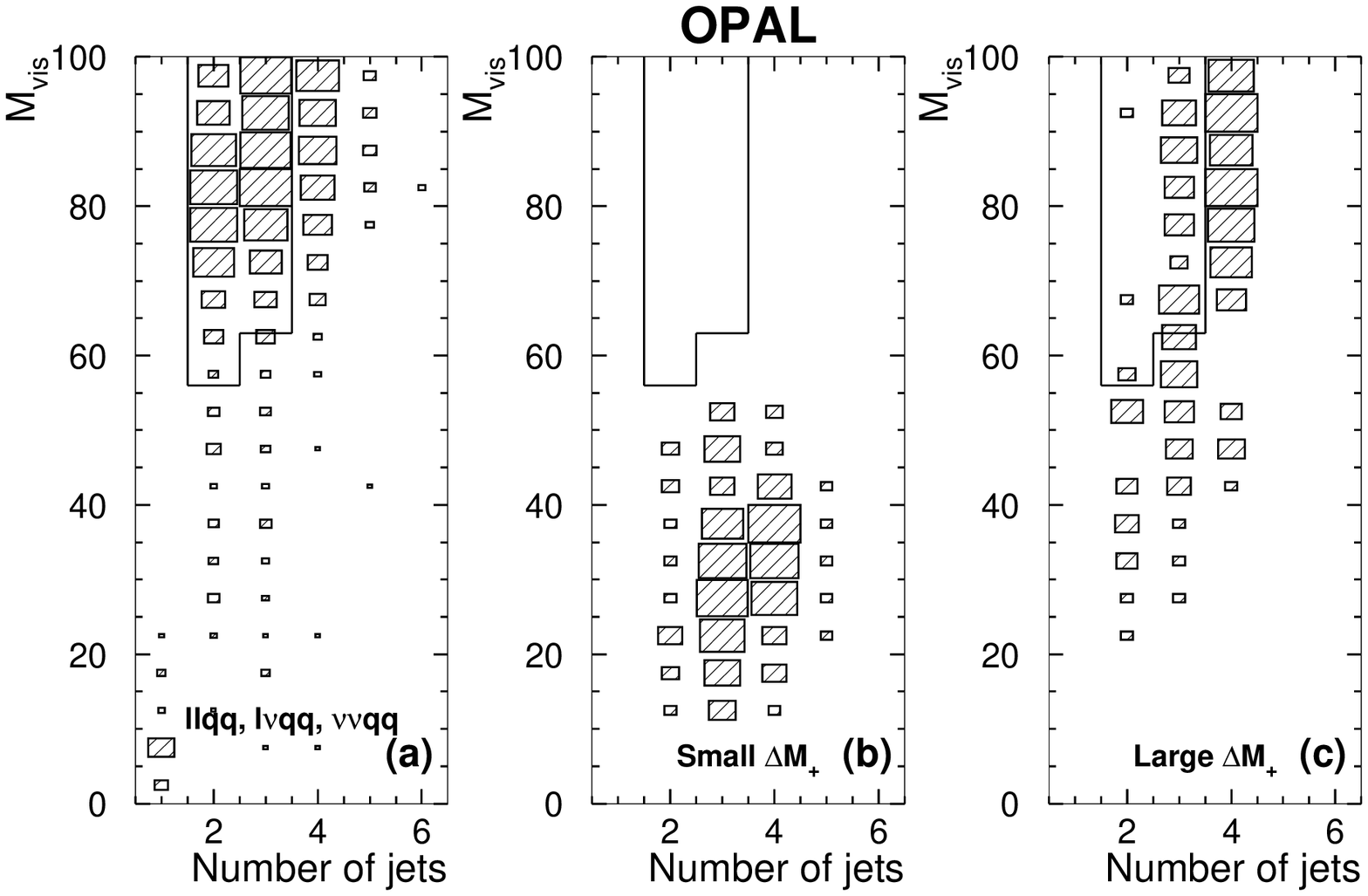,width=16.0cm}
}\end{center}
\caption[1]
{
The distributions of the $\Mvis$ versus $\mathrm{N}_{\mathrm{jets}}$ 
after cut (A4). Distributions of chargino signal events are shown in
(b) for $m_{\ch_1}=80$~GeV and  $m_{\nt_1}=60$~GeV and in (c) for
$m_{\ch_1}=80$~GeV and  $m_{\nt_1}=20$~GeV. 
The background from $\ell\ell\qq$, $\nulqq$ and $\nu\nu\qq$ is shown in
(a). Events with two or three jets are rejected in the region 
outlined by the thin line in the plots. 
}
\label{fig:catamvis}
\end{minipage}
\end{figure}

\begin{figure} \centering
\begin{minipage}{16.0cm}
\begin{center}\mbox{
\epsfig{file=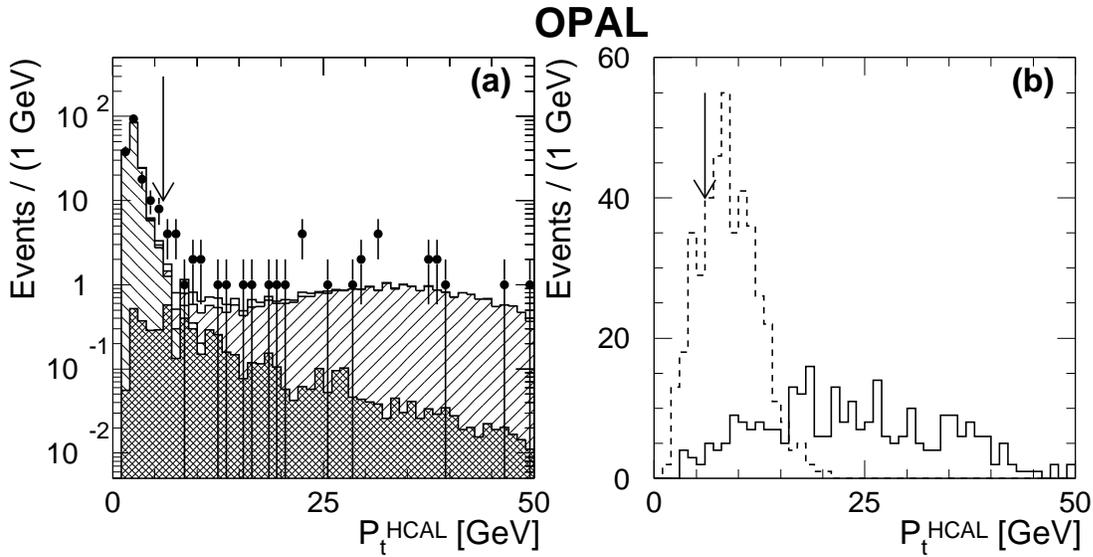,width=16.0cm}}
\end{center}
\vspace{-9mm}
\caption[]
{
The distribution of $P_t^{\mathrm{HCAL}}$ in analysis (B)
after cut (B1).
Data and expected background contributions are shown in (a).
The background sources are labelled as in Fig.~\ref{fig:cataacop}.
The distribution of the signal for simulated chargino events with
$m_{\chp_1} = 80$~GeV and $m_{\nt_1} = 40$~GeV (solid histogram) and
with $m_{\chp_1} = 80$~GeV and $m_{\nt_1} = 70$~GeV (dashed histogram)
are shown in (b). 
The normalizations of the signal distributions are arbitrary. 
}
\label{figb1}
\end{minipage}
\end{figure}

\begin{figure} \centering
\begin{minipage}{16.0cm}
\begin{center}\mbox{
\epsfig{file=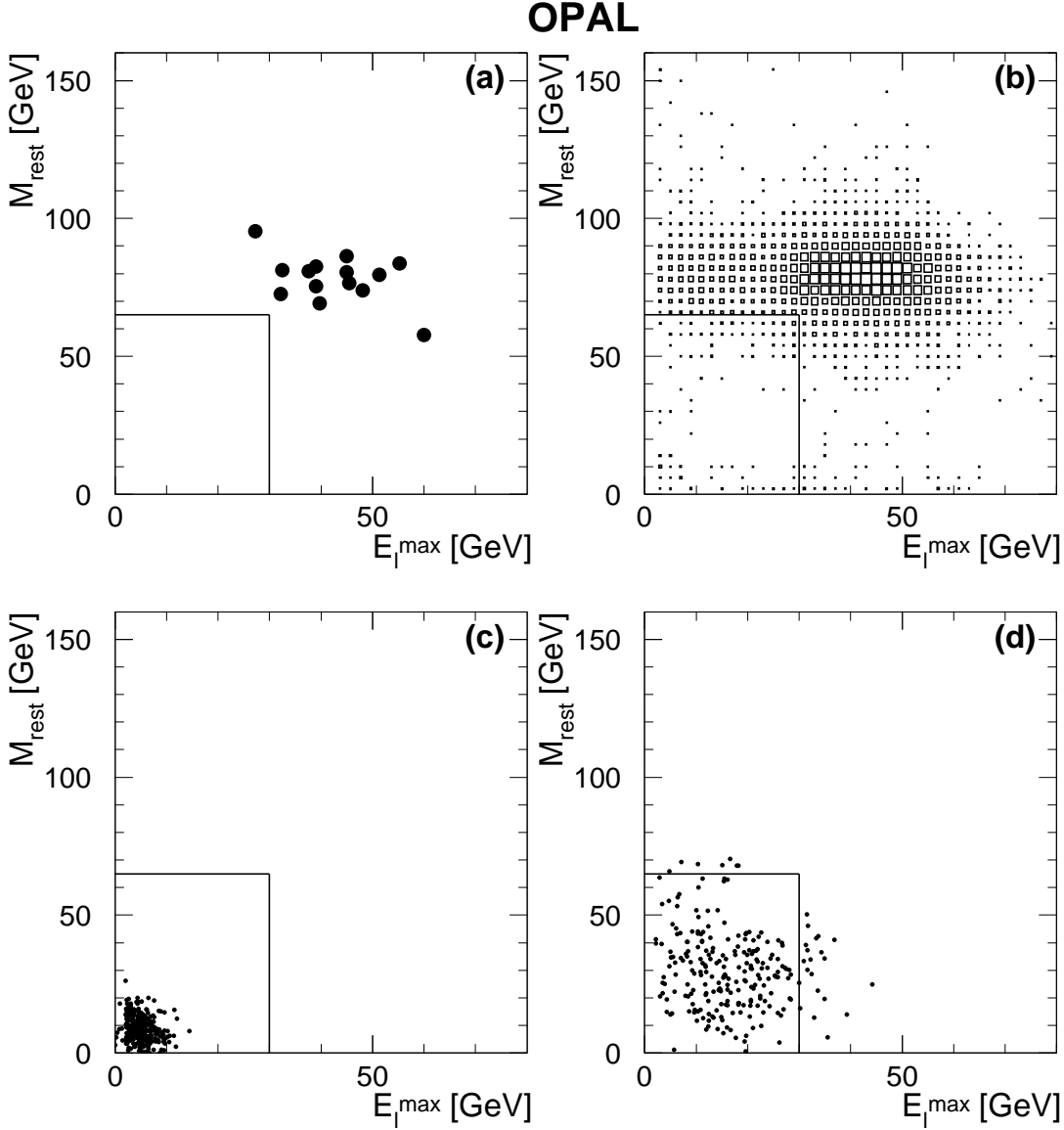,width=16.0cm}}
\end{center}
\vspace{-9mm}
\caption[]
{
Scatter plots of $M_{\mathrm rest}$ vs.\ $E^{\mathrm max}_\ell$ 
after cut (B3).
Figure (a) shows the data. The Monte Carlo prediction for 
four-fermion background is shown in (b). The other figures show 
the distribution of the signal for simulated chargino events with
$m_{\chp_1} = 80$~GeV and $m_{\nt_1} = 70$~GeV (c) and
with $m_{\chp_1} = 80$~GeV and $m_{\nt_1} = 40$~GeV (d).
The straight lines in all figures indicate the selection criteria.
}
\label{figb3}
\end{minipage}
\end{figure}

\begin{figure} \centering
\begin{minipage}{16.0cm}
\begin{center}\mbox{
\epsfig{file=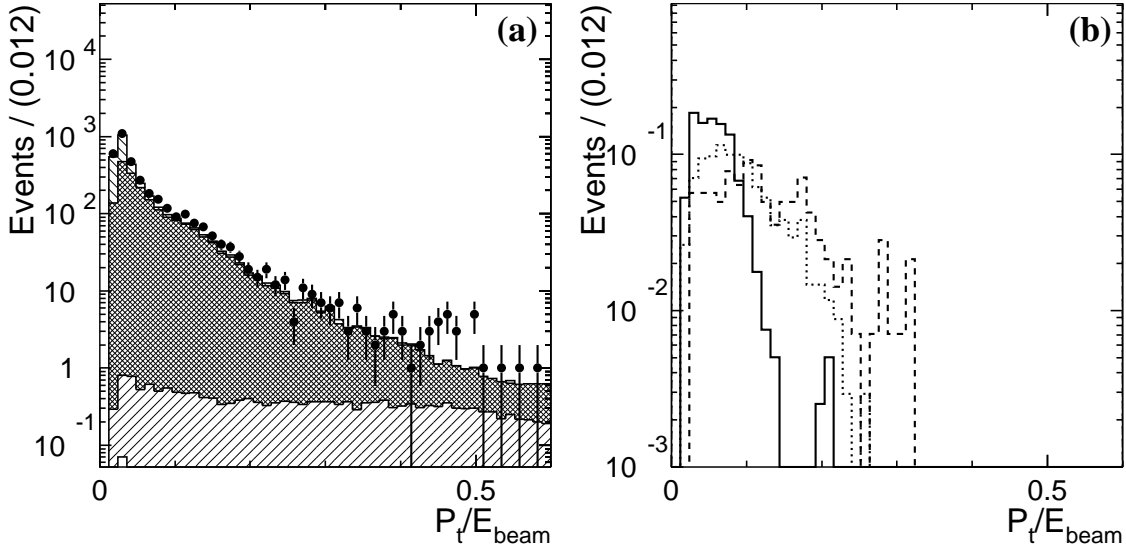,width=16.0cm}}
\end{center}
\vspace{-9mm}
\caption[]
{The distributions of $P_t/E_{\mathrm beam}$ for all $\phi_{\mathrm{acop}}$
after preselection, for events in analysis (C).
Data and background contributions are shown in (a). The
background sources are labelled as in Fig.~\ref{fig:cataacop}.
In (b) predictions from the simulation for chargino
and neutralino events are shown:
$m_{\chp_1}=70$~GeV$,\Delta M_{+}=5$~GeV (solid line),
$m_{\chp_1}=80$~GeV$,\Delta M_{+}=20$~GeV (dashed line),
$m_{\nt_2}+m_{\nt_1}=130$~GeV$,\Delta M_{0}=10$~GeV (dotted line).
The normalizations of the signal distributions are arbitrary. 
}
\label{catc1}
\end{minipage}
\end{figure}

\begin{figure} \centering
\begin{minipage}{16.0cm}
\begin{center}\mbox{
\epsfig{file=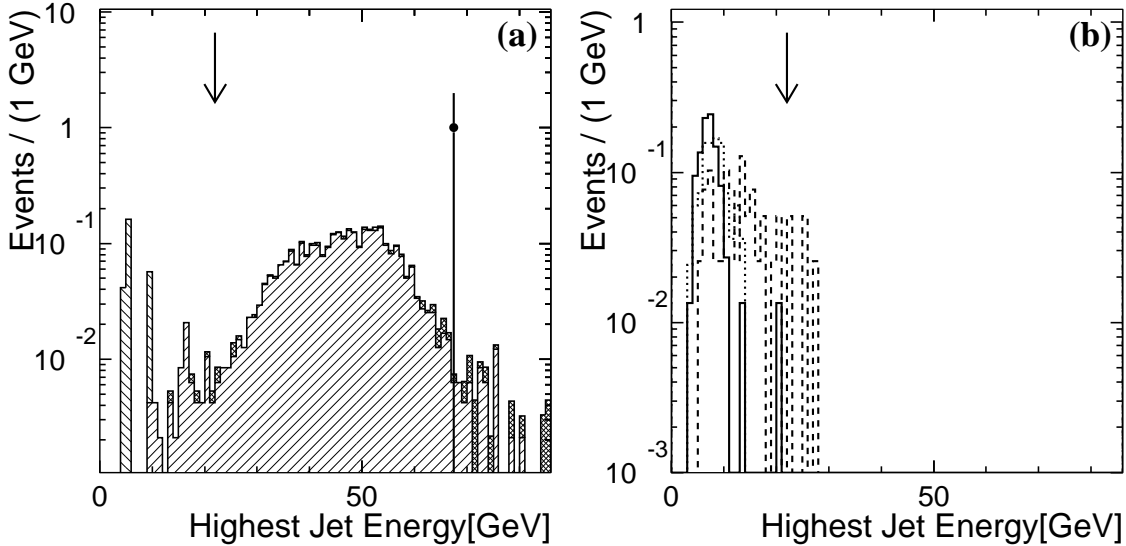,width=16.0cm}}
\end{center}
\vspace{-9mm}
\caption[]
{
The distributions of the jet energy of the events
in category (C) after all the other cuts were applied.
Data and background contributions are shown in (a). The
background sources are labelled as in Fig.~\ref{fig:cataacop}.
The Monte Carlo signal distributions are shown in (b) and 
are labelled as in Fig.~\ref{catc1}.
The normalizations of the signal distributions are arbitrary. 
}
\label{catc3}
\end{minipage}
\end{figure}

\begin{figure} \centering
\begin{minipage}{16.0cm}
\begin{center}\mbox{
\epsfig{file=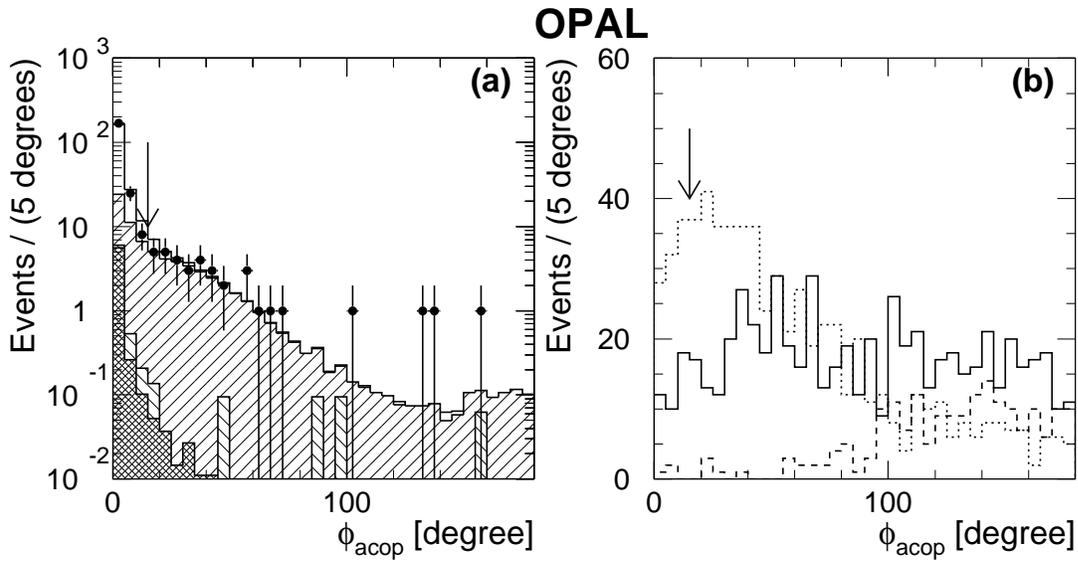,width=16.0cm}}
\end{center}
\vspace{-9mm}
\caption[]
{
The distribution of the acoplanarity angle
in analysis (D) after cut (D2)\@.
Data and background prediction from the simulation are shown in (a). The
background sources are labelled as in Fig.~\ref{fig:cataacop}.
In (b) Monte Carlo predictions from
$\nt_1 \nt_2$ events are given for 
$m_{\nt_2} =  95$~GeV and $m_{\nt_1} = 65$~GeV (solid line),
$m_{\nt_2} =  85$~GeV and $m_{\nt_1} = 75$~GeV (dashed line) and
$m_{\nt_2} = 115$~GeV and $m_{\nt_1} = 45$~GeV (dotted line).
The normalizations of the signal distributions are arbitrary. 
}
\label{figdacop}
\end{minipage}
\end{figure}

\begin{figure} \centering
\begin{minipage}{16.0cm}
\begin{center}\mbox{
\epsfig{file=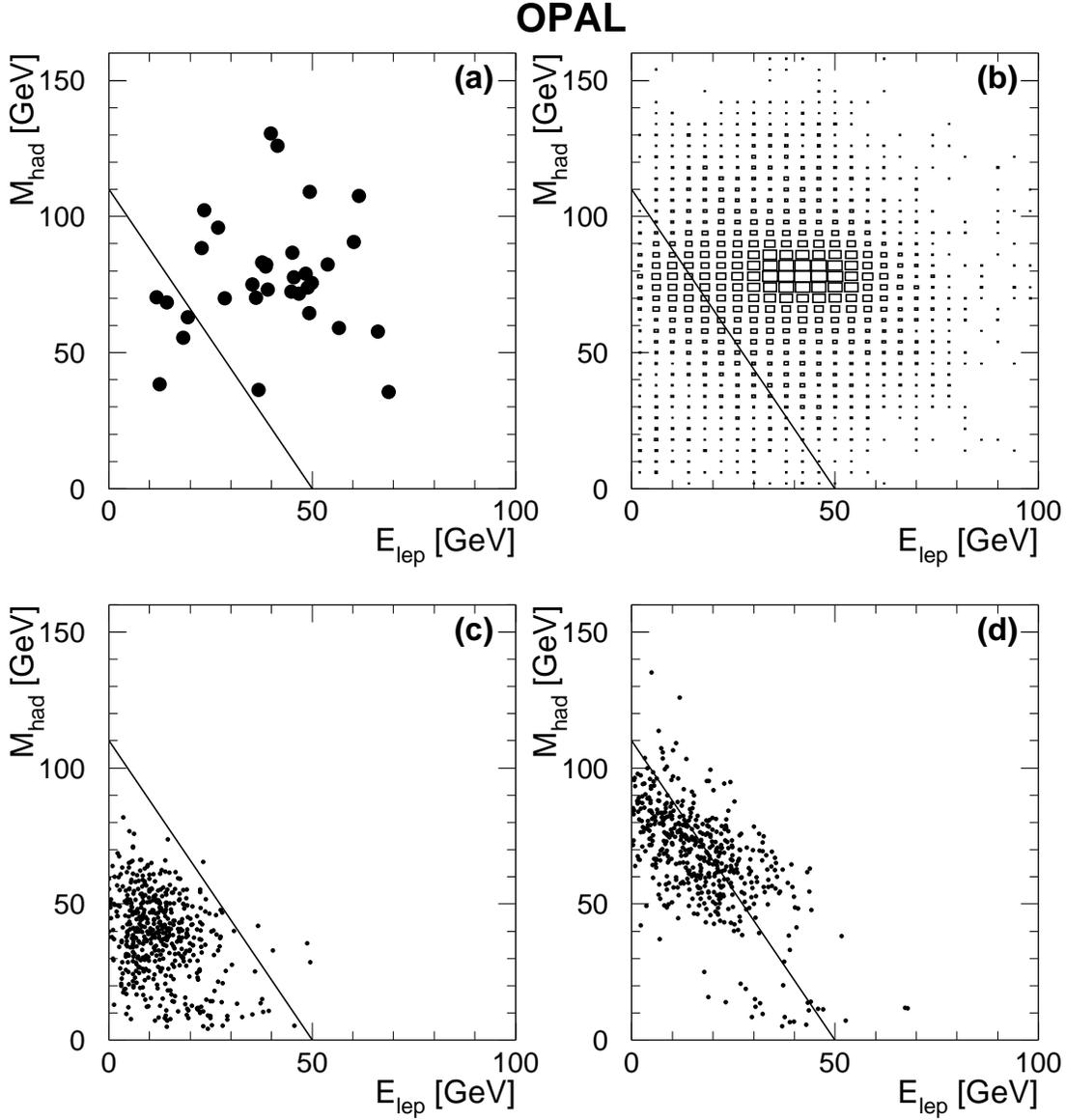,width=16.0cm}}
\end{center}
\vspace{-9mm}
\caption[]
{
Scatter plots of $ {\mathrm M_{had}} $ versus
$ {\mathrm E_{lep}} $ in analysis (D) after cut (D3)
for those events whose  ${\mathrm M_{vis}}$ is larger than 20~GeV\@.
The data are shown in (a). Figure (b) shows the Monte Carlo prediction
for the four-fermion
background. In figures (c) and (d) signal distributions from
simulated $\nt_1 \nt_2$ events are given for 
$m_{\nt_2} = 115$~GeV and $m_{\nt_1} = 45$~GeV (c) and
$m_{\nt_2} = 135$~GeV and $m_{\nt_1} = 25$~GeV (d).
The straight lines in these figures indicate the selection cut.
}
\label{figdmhad}
\end{minipage}
\end{figure}

\begin{figure} \centering
\begin{minipage}{16.0cm}
\begin{center}\mbox{
\epsfig{file=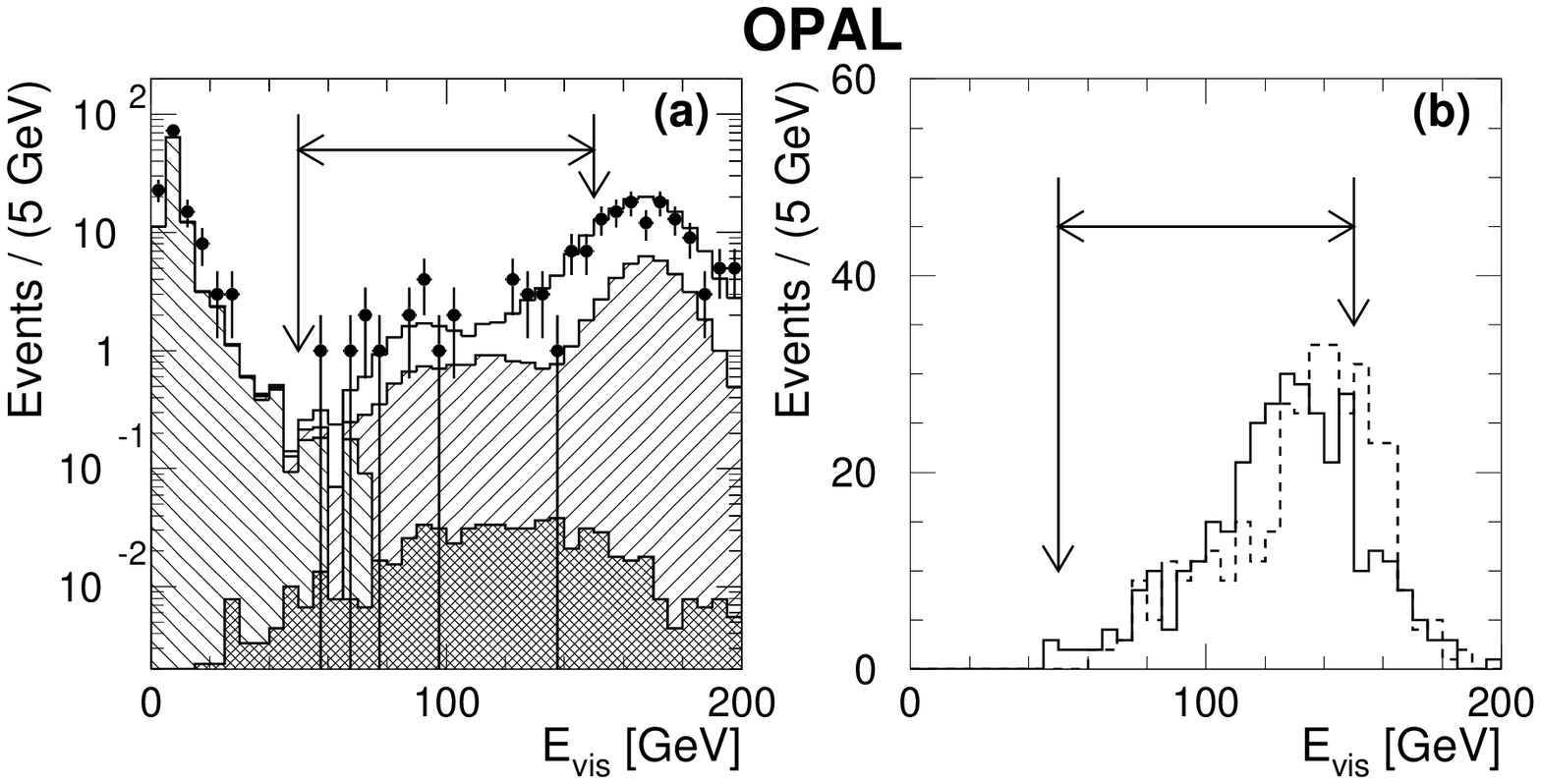,width=16.0cm}
}
\end{center}
\vspace{-9mm}
\caption[]
{
The distribution of $E_{\mathrm vis}$ in analysis (E)
after cut (E3).
Data and expected Monte Carlo background contributions are shown in (a).
The background sources are labelled as in Fig.~\ref{fig:cataacop}.
The distributions of the signal for simulated chargino events with
$m_{\chp_1} = 80$~GeV and $m_{\nt_1} = 0.6$~GeV 
with the $\chp_1 \ra \nt_1 \Wrv$ decay (solid histogram) and
$m_{\chp_1} = 80$~GeV, $m_{\nt_2} = 2$~GeV and $m_{\nt_1} = 0.6$~GeV
with the cascade decay $\chp_1 \ra \nt_2 \Wrv \ra \nt_1 \gamma \Wrv$ 
(dashed histogram) are shown in (b).
The normalizations of the signal distributions are arbitrary. 
}
\label{fige1}
\end{minipage}
\end{figure}

\begin{figure} \centering
\begin{minipage}{16.0cm}
\begin{center}\mbox{
\epsfig{file=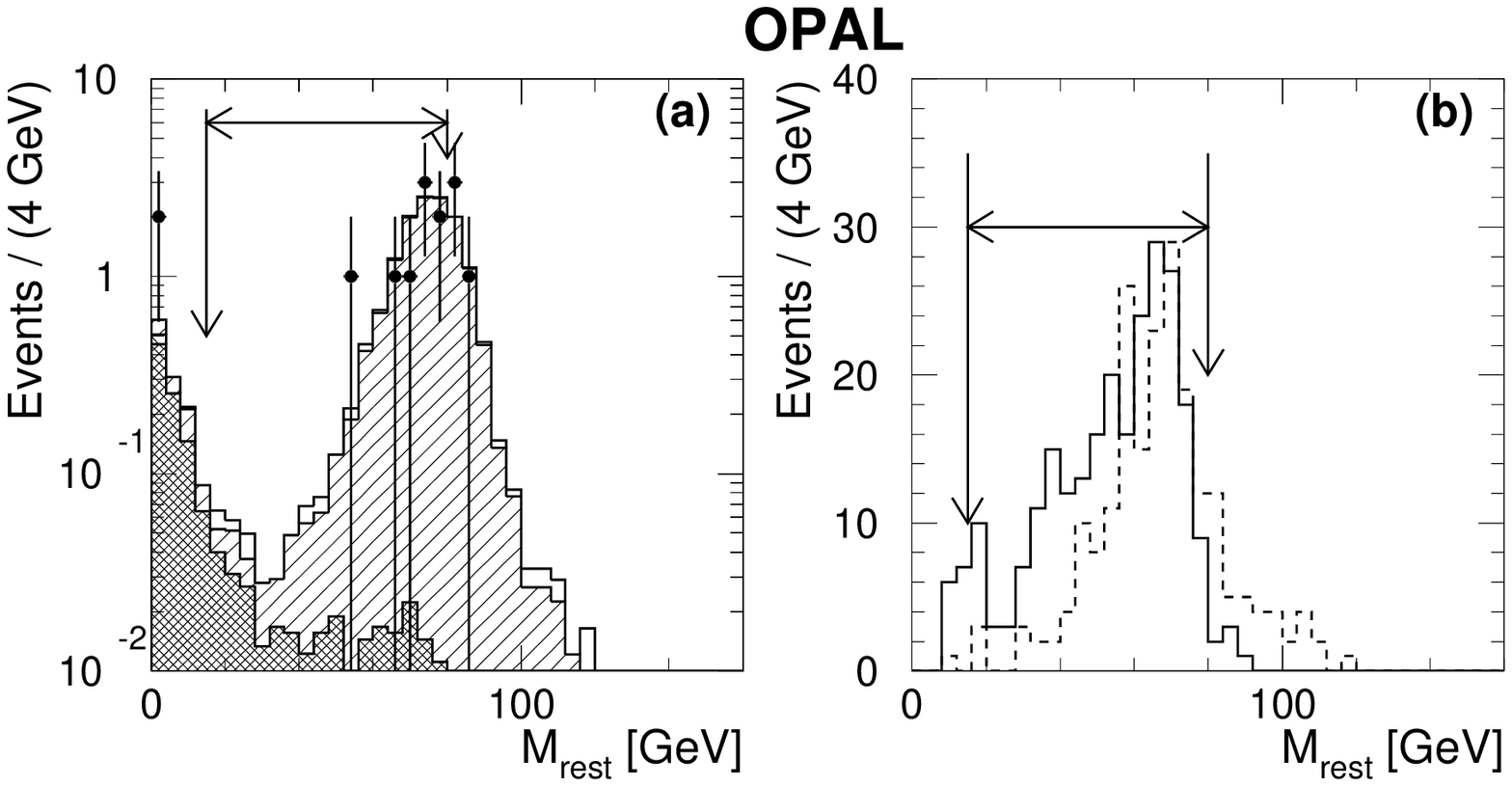,width=16.0cm}
}
\end{center}
\vspace{-9mm}
\caption[]
{
The distribution of $M_{\mathrm rest}$ in analysis (F)
after cut (F4).
Data and expected Monte Carlo background contributions are shown in (a).
The background sources are labelled as in Fig.~\ref{fig:cataacop}.
The distributions of the signal for simulated chargino events with
$m_{\chp_1} = 80$~GeV and $m_{\nt_1} = 0.6$~GeV 
with the $\chp_1 \ra \nt_1 \Wrv$ decay (solid histogram) and
$m_{\chp_1} = 80$~GeV, $m_{\nt_2} = 2$~GeV and $m_{\nt_1} = 0.6$~GeV
with the cascade decay $\chp_1 \ra \nt_2 \Wrv \ra \nt_1 \gamma \Wrv$ 
(dashed histogram) are shown in (b).
The normalizations of the signal distributions are arbitrary. 
}
\label{figf1}
\end{minipage}
\end{figure}

\begin{figure} \centering
\begin{minipage}{16.0cm}
\begin{center}\mbox{
\epsfig{file=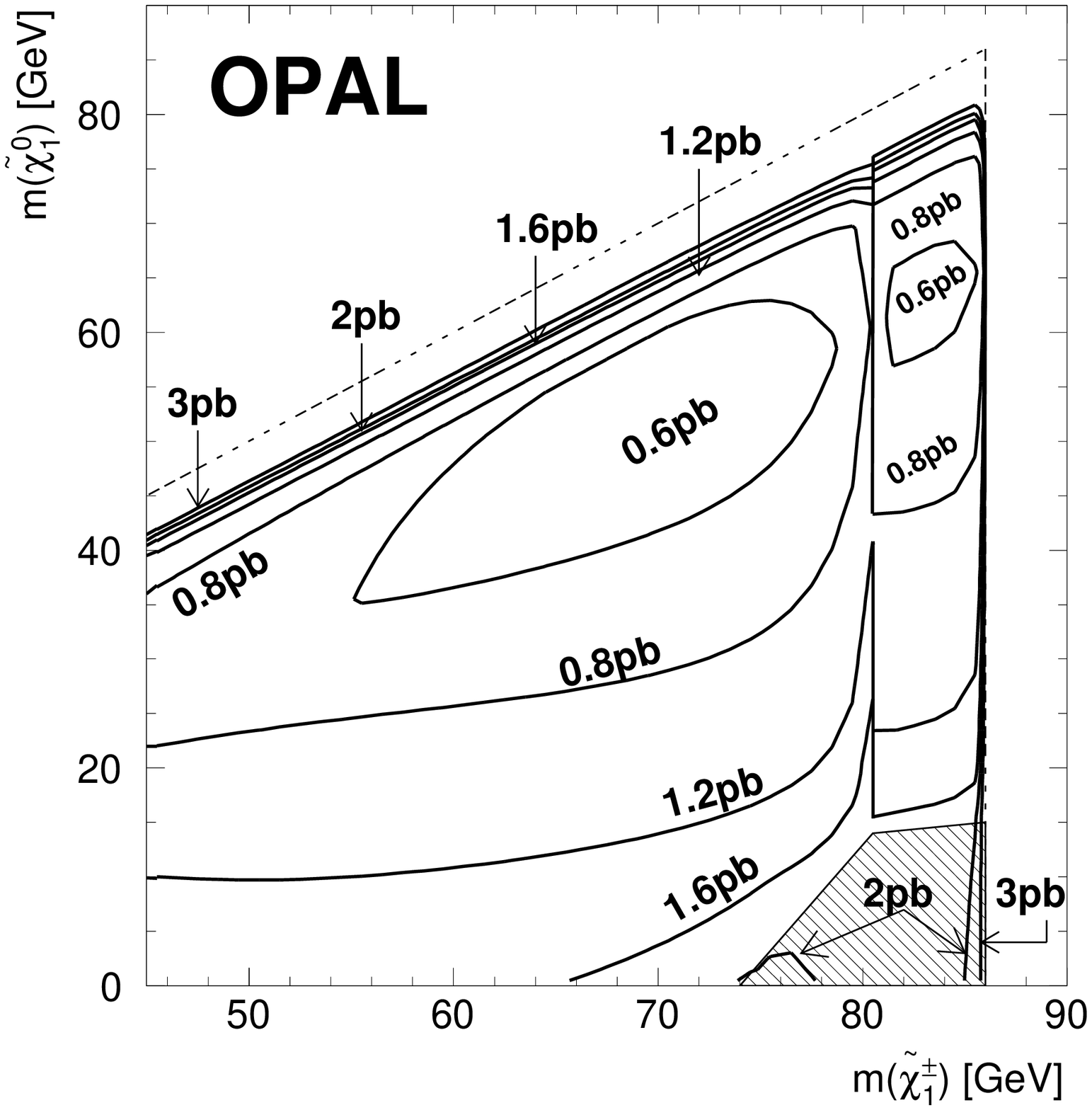,width=16.0cm}}
\end{center}
\vspace{-9mm}
\caption[]
{
The contours of the 95\% C.L. upper limits
for the $\ee \ra \chp_1 \chm_1$ production cross-sections at
$\roots =$ 172~GeV are shown
assuming Br$(\chp_1 \ra \nt_1 \Wvp)=100$\%. 
These limits have been obtained by combining the results
from 161, 170 and 172~GeV assuming that the cross-sections scale with
$\tilde \beta /s$, as described in the text.
The hatched area indicates the region where analyses
(E) and (F) contribute.
}
\label{figsum1}
\end{minipage}
\end{figure}

\begin{figure} \centering
\begin{minipage}{16.0cm}
\begin{center}\mbox{
\epsfig{file=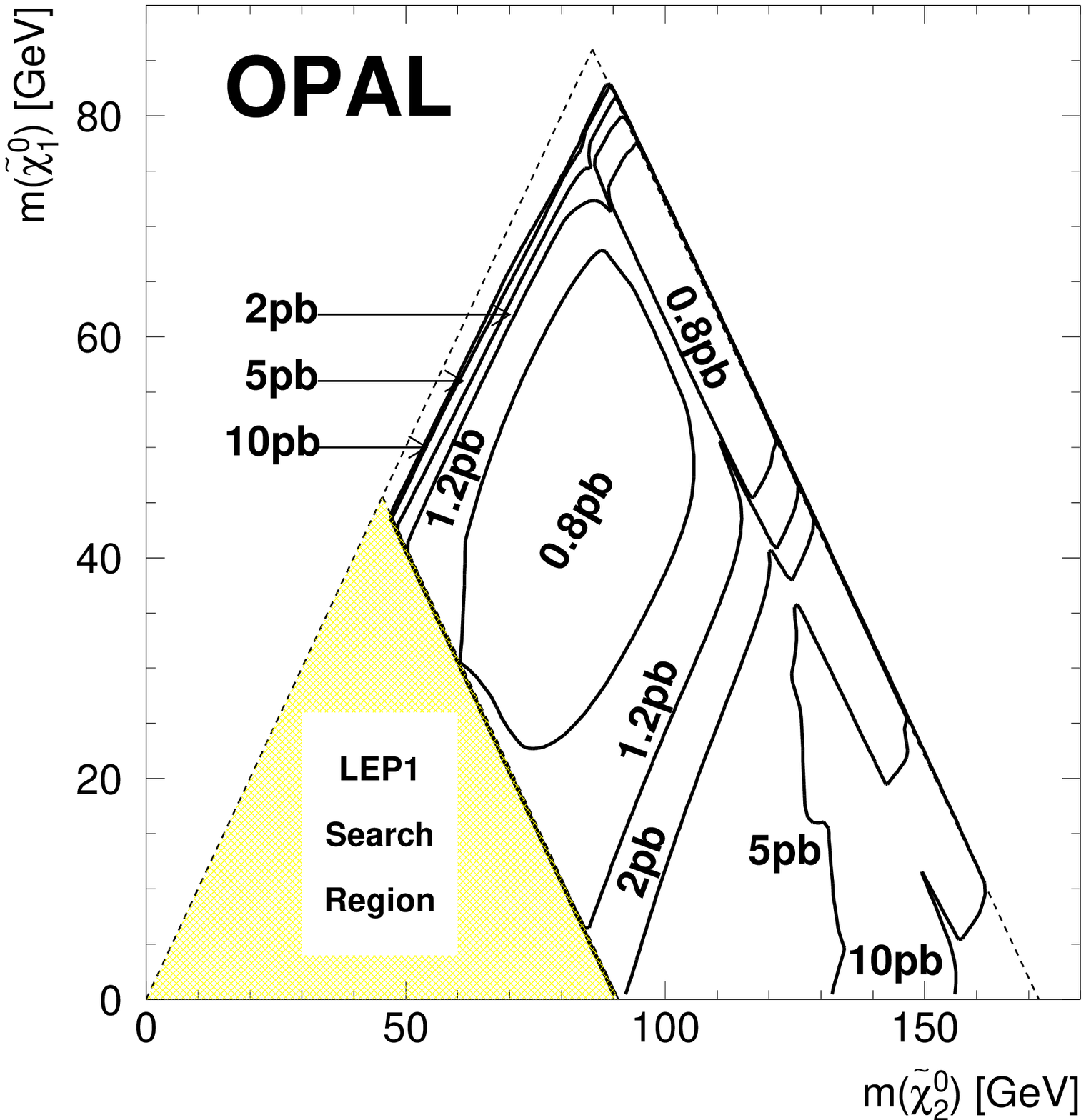,width=16.0cm}}
\end{center}
\vspace{-9mm}
\caption[]
{
The contours of the 95\% C.L. upper limits for
the $\ee \ra \nt_2 \nt_1$ production cross-sections at
$\roots = $172~GeV are shown assuming 
Br$(\nt_2 \ra \nt_1 \Zv) = 100$\%.
The region for which 
$(m_{\nt_2} + m_{\nt_1}) < m_{\mathrm Z}$ is not
considered in this analysis.
These limits have been obtained by combining the results
from 161, 170 and 172~GeV assuming that the cross-sections scale with
$\sqrt{\tilde\beta_1 \cdot \tilde\beta_2} /s$, as described in the text.
}
\label{figsum2}
\end{minipage}
\end{figure}

\begin{figure} \centering
\begin{minipage}{16.0cm}
\begin{center}\mbox{
\epsfig{file=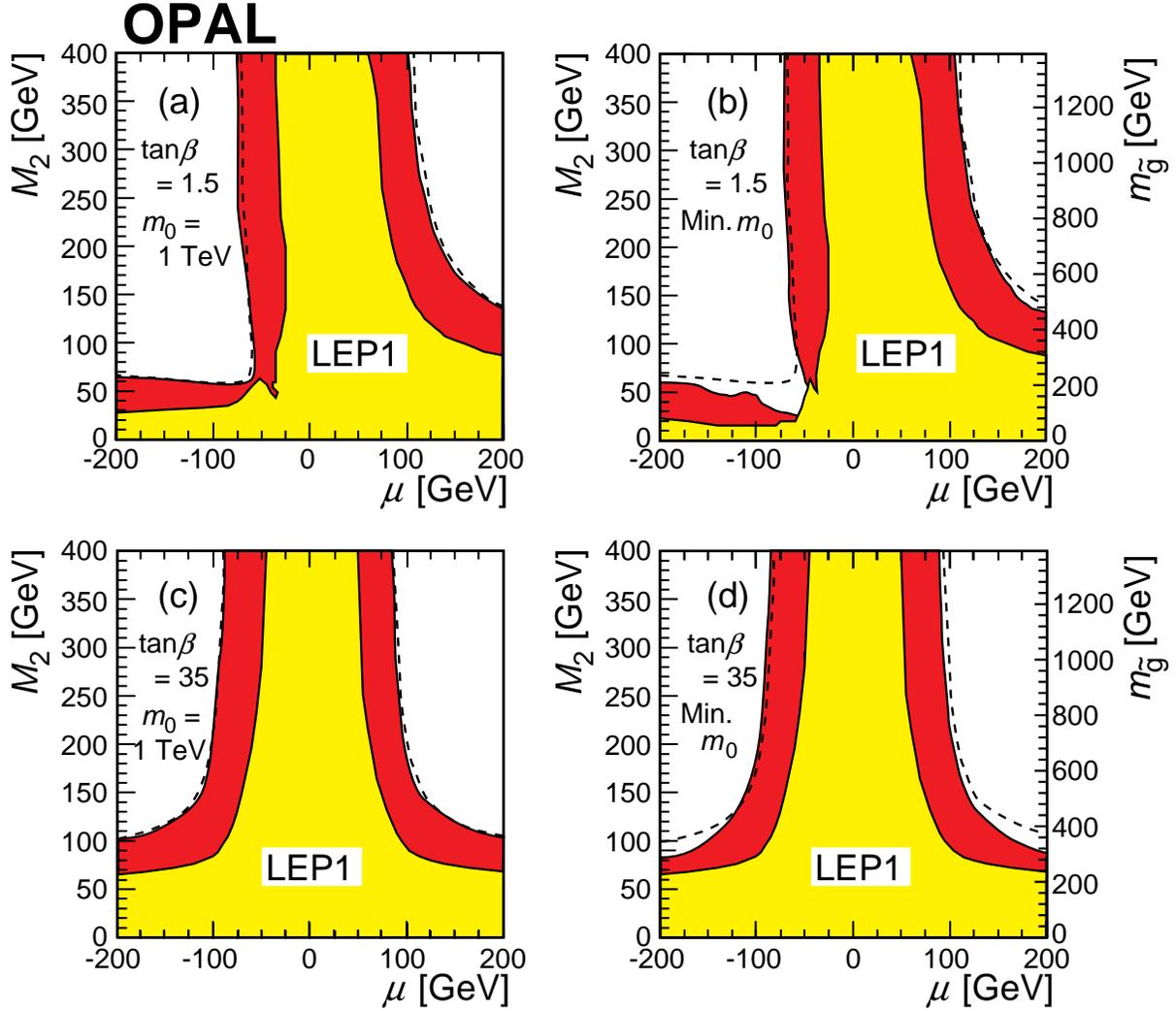,width=16.0cm}}
\end{center}
\vspace{-9mm}
\caption[]
{
Exclusion regions at 95\% C.L. in the ($M_2$,$\mu$) plane
for (a) $\tan \beta=1.5$ and $m_0=1$~TeV, (b) $\tan \beta=1.5$ and the
minimum $m_0$ case (see text for details), (c) $\tan \beta=35$ and $m_0=1$~TeV,
and (d) $\tan \beta=35$ and
the minimum $m_0$ case.
The light shaded areas show the LEP1 excluded regions
and the dark shaded areas show the additional excluded
region using the data from $\sqrt{s}=161$~GeV combined with those
from 170 and 172~GeV.
The kinematical boundary for $\chp_1 \chm_1$ production is shown
by the dashed curves. 
The corresponding mass limits on the gluino mass, $m_{\tilde{g}}$,
in the CMSSM are also shown.
}
\label{fig_mssm}
\end{minipage}
\end{figure}

\begin{figure} \centering
\begin{minipage}{16.0cm}
\vspace{-4cm}
\begin{center}\mbox{
\epsfig{file=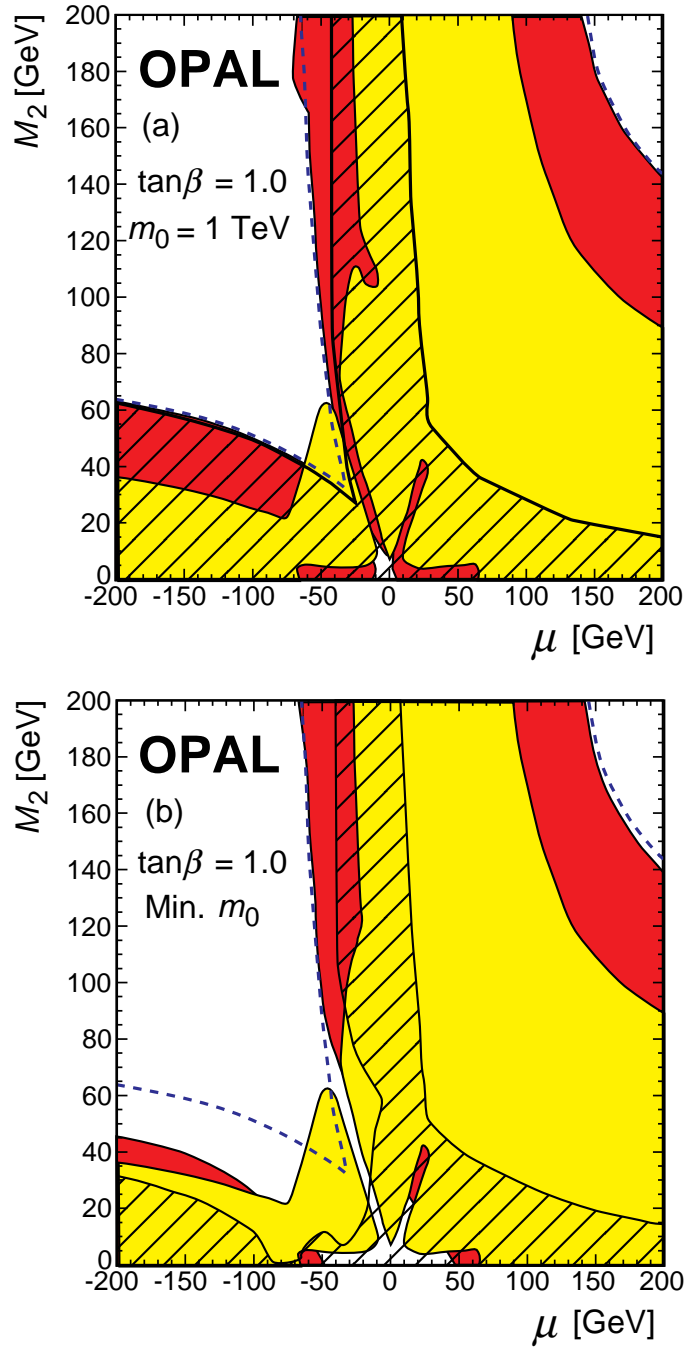,width=9.0cm}}
\end{center}
\vspace{-9mm}
\caption[]
{
Exclusion regions at 95\% C.L. in the ($M_2$,$\mu$) plane
for (a) $\tan \beta=1.0$ and $m_0=1$~TeV, and (b) for the
the minimum $m_0$ case.
The light shaded areas show the LEP1 excluded regions~\cite{LEP1-neut},
and the dark shaded areas show the additional excluded
region using analyses (A)-(D) and the data from $\sqrt{s}=161-172$~GeV.
The hatched area shows the excluded region obtained with analyses (E) and
(F) and the data at $\sqrt{s}=170-172$~GeV.
The kinematical boundary for $\chp_1 \chm_1$ production is shown
by the dashed curves. 
}
\label{label_12}
\end{minipage}
\end{figure}

\begin{figure} \centering
\begin{minipage}{16.0cm}
\begin{center}\mbox{
\epsfig{file=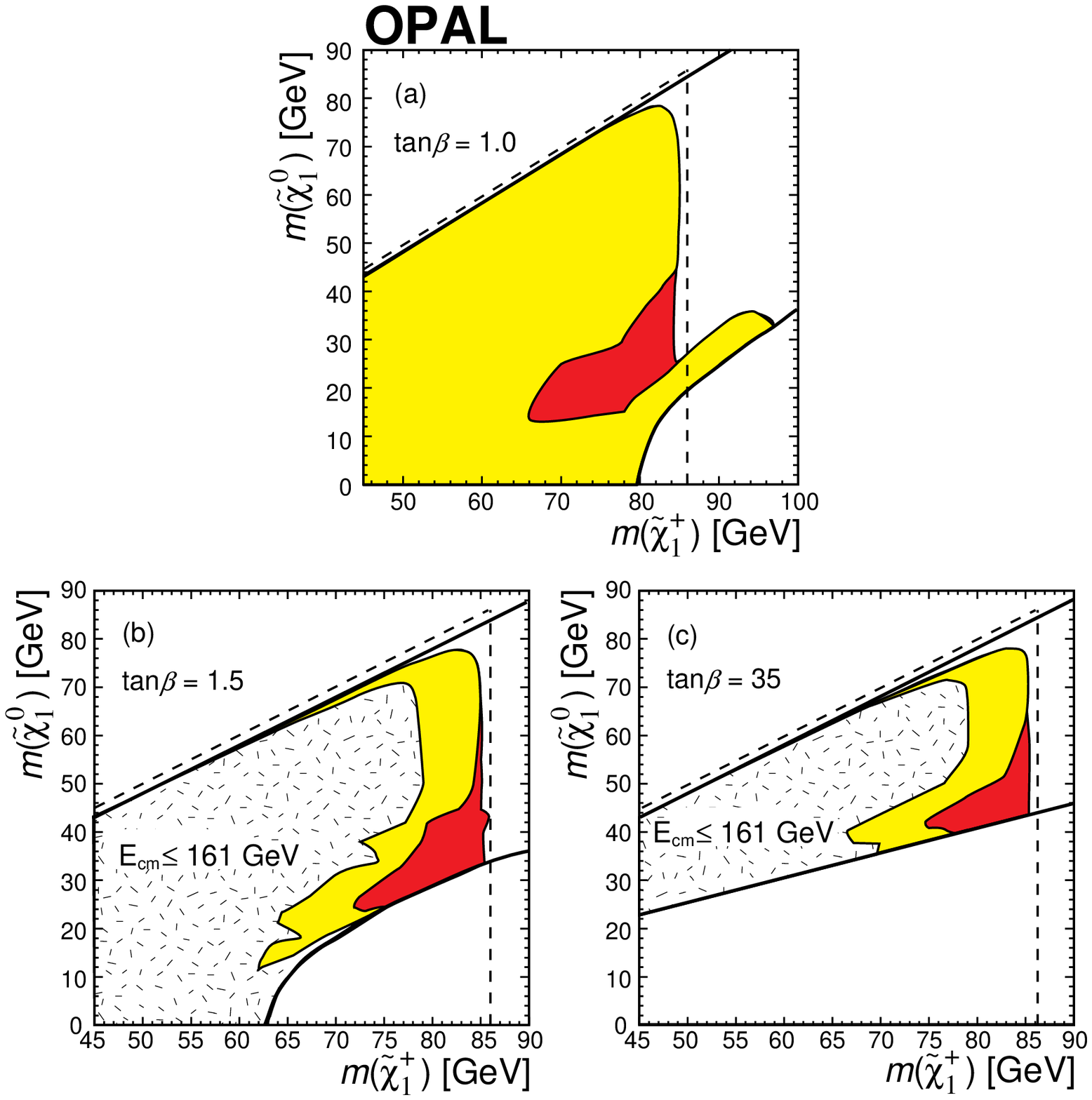,width=16.0cm}
}
\end{center}
\vspace{-9mm}
\caption[]
{
The 95\% C.L. excluded region
in the ($m_{\chp_1}$,$m_{\nt_1}$) plane
within the framework of the MSSM
for the case of minimum $m_0$ (light shaded region) and
$m_0 = 1$~TeV (including also dark shaded region) for
(a) $\tan \beta=1.0$, (b) $\tan \beta = 1.5$ and
(c) $\tan \beta = 35$.
The region excluded by the analysis of LEP1, LEP1.5~\cite{LEP15-opal}
and 161~GeV~\cite{LEP16-opal} data
is also shown in (b) and (c) for the minimum $m_0$ case
(speckled region).
The thick solid lines represent the theoretical bounds
of the MSSM parameter space as given in the text.
The kinematical boundaries for $\chp_1 \chm_1$ production and
decay at $\roots = 172$~GeV are shown by
dashed lines.
}
\label{masslimc}
\end{minipage}
\end{figure}

\begin{figure} \centering
\begin{minipage}{16.0cm}
\begin{center}\mbox{
\epsfig{file=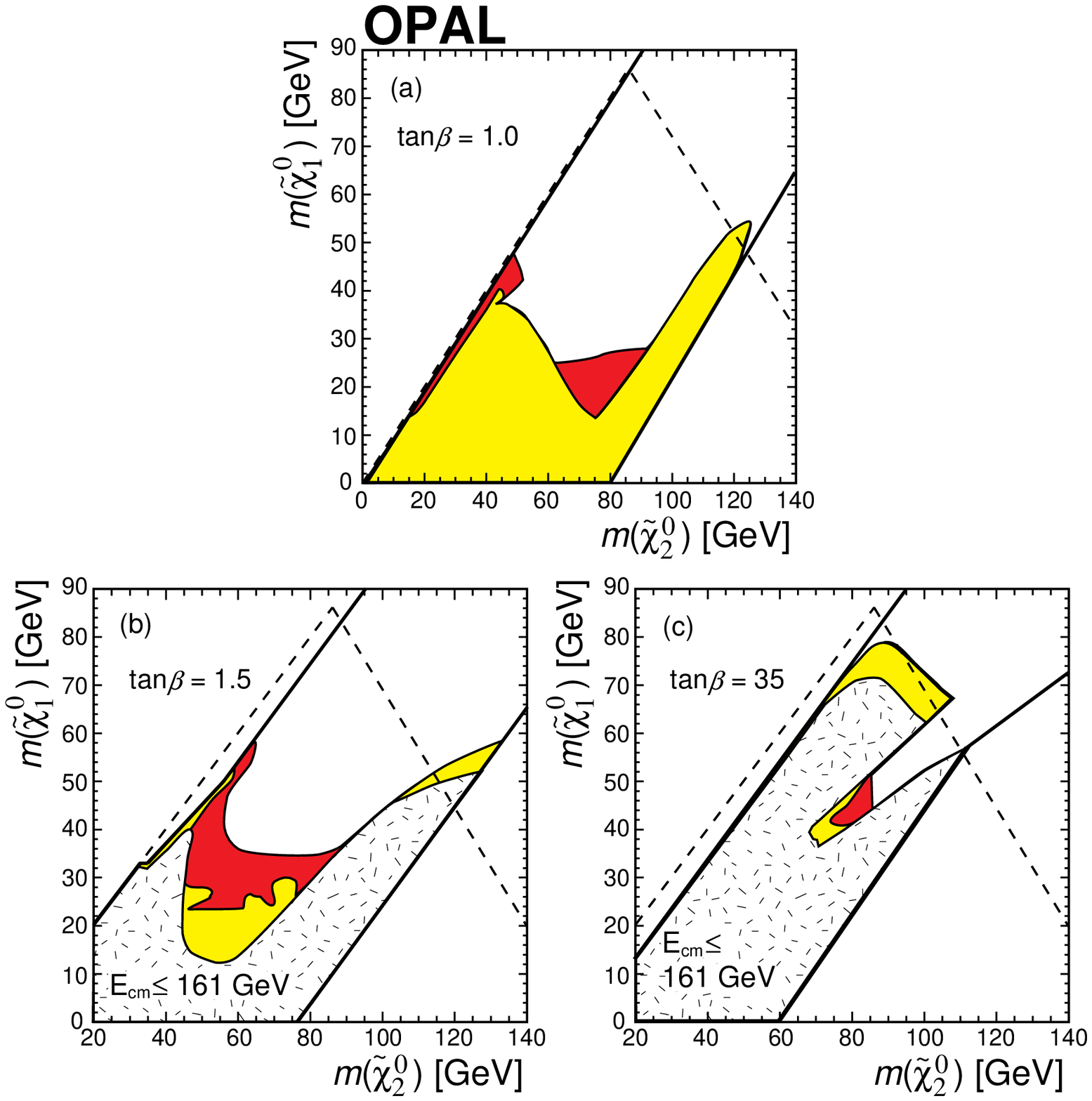,width=16.0cm}
}
\end{center}
\vspace{-9mm}
\caption[]
{
The 95\% C.L. excluded region
in the ($m_{\nt_2}$,$m_{\nt_1}$) plane
within the framework of the MSSM
for the case of minimum $m_0$ (light shaded region) and
$m_0 = 1$~TeV (including also dark shaded region) for
(a) $\tan \beta=1.0$, (b) $\tan \beta = 1.5$ and
(c) $\tan \beta = 35$.
The region excluded by the analysis of LEP1, LEP1.5~\cite{LEP15-opal}
and 161~GeV~\cite{LEP16-opal} data
is also shown in (b) and (c) for the minimum $m_0$ case
(speckled region).
The thick solid lines represent the theoretical bounds
of the MSSM parameter space as given in the text.
The kinematical boundaries for $\nt_1 \nt_2$ production and
decay at $\roots = 172$~GeV are shown by
dashed lines.
}
\label{masslimn}
\end{minipage}
\end{figure}

\begin{figure} \centering
\begin{minipage}{16.0cm}
\begin{center}\mbox{
\epsfig{file=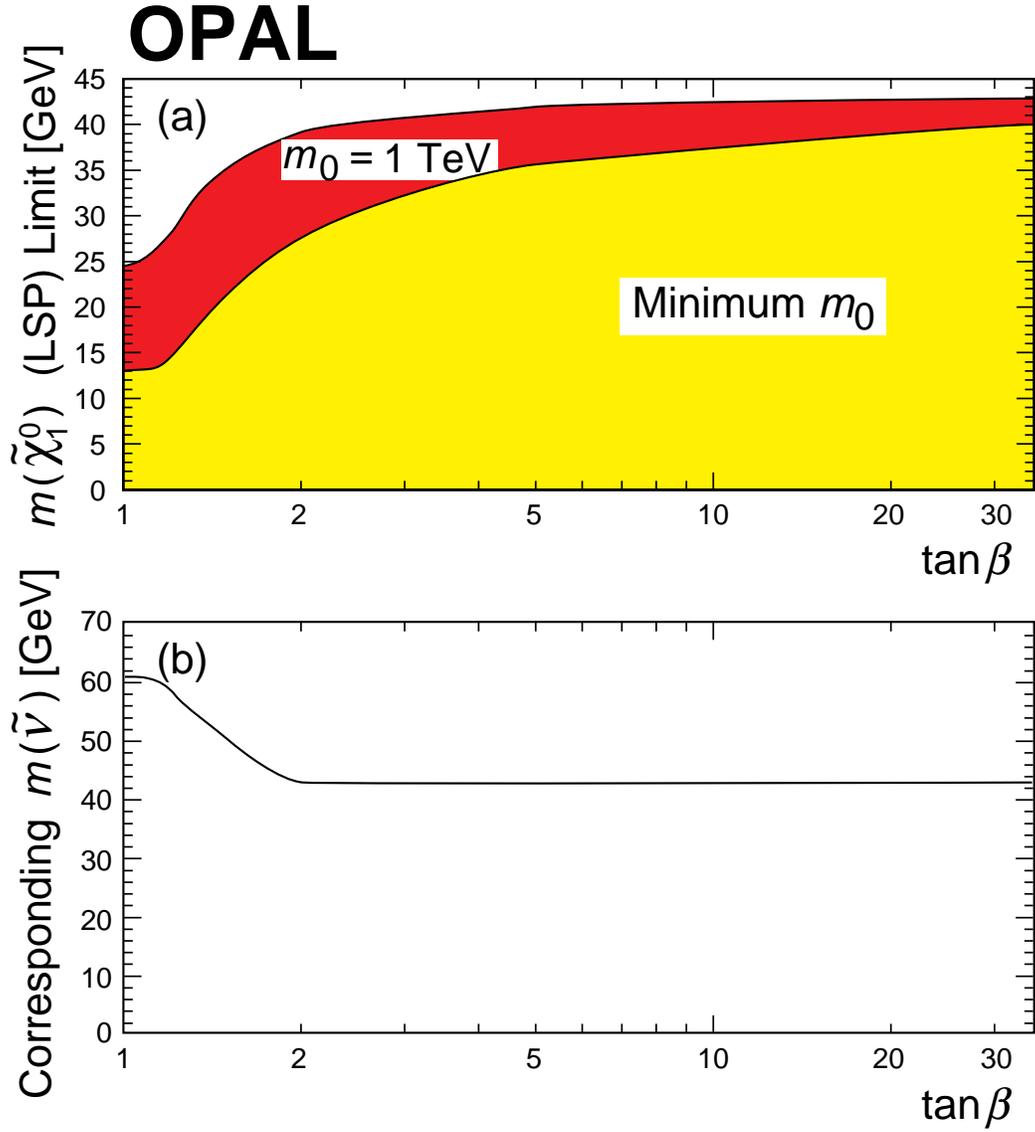,width=14.0cm}
}
\end{center}
\vspace{-9mm}
\caption[]
{
(a) The lower mass limit on the mass of the $\nt_1$ as a function
of $\tan\beta$ for $m_0 = 1$~TeV and minimum $m_0$.
(b) The
value of $m_{\tilde{\nu}}$ corresponding to the minimum $m_0$ value
used as a function of $\tan\beta$.
}
\label{mass_tanb}
\end{minipage}
\end{figure}

\begin{figure} \centering
\begin{minipage}{16.0cm}
\vspace{-2cm}
\begin{center}\mbox{
\epsfig{file=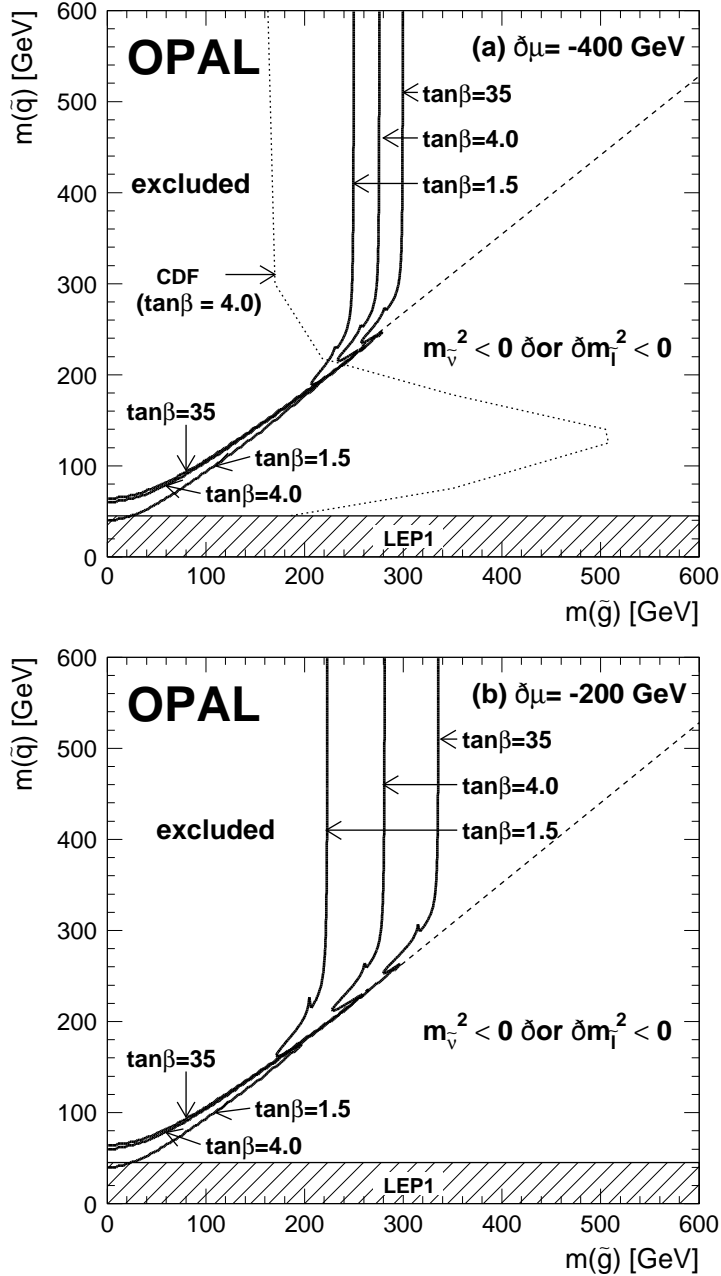,width=10.0cm}}
\end{center}
\vspace{-9mm}
\caption[]
{
Limits on $M_2$ interpreted as limits on $m_{\sg}$ assuming 
gauge unification at the GUT scale by using the relation 
$m_{\tilde{g}}/M_2=\alpha_s/\frac{\alpha}{{\mathrm sin}^2 \theta_W}=3.5$.
$m_{\sq}$ is the average of the $\su_R$, $\su_L$, $\sd_R$ and $\sd_L$ 
masses and can be calculated from $m_0$, $M_2$ and $\tan \beta$ 
in the CMSSM framework.
The limits in the ($m_{\sq}$,$m_{\sg}$) plane 
for (a) $\mu = -400$~GeV and (b) $\mu = -200$~GeV
thus obtained are shown.
The limit in (a) can be compared with the current direct search mass limits 
on $\sq$ and $\sg$ from the CDF experiment~\cite{CDFsgsq}.
The hatched region labelled LEP1 has been excluded by the $\Z$ width
measurement at LEP1~\cite{PDG}.
In the region below the dashed diagonal line the lightest slepton 
or sneutrino mass becomes negative within the framework of 
the CMSSM.
}
\label{squark1}
\end{minipage}
\end{figure}

\end{document}